%% file: DESY-07-052.tex
\documentclass[zpreprint,zbstnp]{zeus_paper}
\usepackage[english]{babel}
\usepackage{lineno}
\usepackage{lscape}
\input zeus_def.tex

\input charm_com.tex
\input physics_com.tex
\def\citeCTD{{\cite{%
nim:a279:290,*npps:b32:181,*nim:a338:254%
}}\xspace}
\def\citeCAL{{\cite{%
nim:a309:77,*nim:a309:101,*nim:a321:356,*nim:a336:23%
}}\xspace}

\newcommand{\ftwo}      {\mbox{$F_{2}$}}
\newcommand{\ftwoccb}   {\mbox{$F_{2}^{c\bar c}$}}
\begin{document}
\prepnum{DESY--07--052}

\title{
Measurement of D mesons Production \\
in Deep Inelastic Scattering at HERA
}

\author{ZEUS Collaboration}
\date{April 2007}

\abstract{

Charm production in deep inelastic scattering has been measured with the ZEUS
detector at HERA using an integrated luminosity of 82~pb$^{-1}$.
Charm has been tagged by reconstructing $D^{*+}$, $D^0$, $D^{+}$ and  $D_s^+$
(+ c.c.) charm mesons. The charm hadrons were measured in the kinematic range
$p_T(D^{*+},D^0,D^{+}) > 3$ GeV, $p_T(D_s^+)>2$ GeV and $|\eta(D)| < 1.6$
for $1.5 < Q^2 < 1000$ GeV$^2$ and $0.02 < y < 0.7$.
The production cross sections were used to extract charm fragmentation ratios
and the fraction of $c$ quarks hadronising into a particular charm
meson in the kinematic range considered. The cross sections were
compared to the predictions of next-to-leading-order QCD, and
extrapolated to the full kinematic region in $p_T(D)$ and $\eta(D)$ in order to
determine the open-charm contribution, $F_2^{c\bar{c}}(x,Q^2)$, to the proton
structure function $F_2$.
}

\makezeustitle
\def\3{\ss}
\pagenumbering{Roman}
%
                                                   %
\begin{center}
{                      \Large  The ZEUS Collaboration              }
\end{center}
  S.~Chekanov$^{   1}$,
  M.~Derrick,
  S.~Magill,
  B.~Musgrave,
  D.~Nicholass$^{   2}$,
  \mbox{J.~Repond},
  R.~Yoshida\\
 {\it Argonne National Laboratory, Argonne, Illinois 60439-4815}, USA~$^{n}$
\par \filbreak
  M.C.K.~Mattingly \\
 {\it Andrews University, Berrien Springs, Michigan 49104-0380}, USA
\par \filbreak
  M.~Jechow, N.~Pavel~$^{\dagger}$, A.G.~Yag\"ues Molina \\
  {\it Institut f\"ur Physik der Humboldt-Universit\"at zu Berlin,
           Berlin, Germany}
% ...
\newpage           
$^{\    1}$ supported by DESY, Germany \\
$^{\    2}$ also affiliated with University College London, UK \\
% ...
$^{\dagger}$ deceased \\
%
                                                           %
                                                           %
\begin{tabular}[h]{rp{14cm}}
$^{a}$ &  supported by the Natural Sciences and Engineering Research Council of Canada (NSERC) \\
$^{b}$ &  supported by the German Federal Ministry for Education and Research (BMBF), under
          contract numbers HZ1GUA 2, HZ1GUB 0, HZ1PDA 5, HZ1VFA 5\\
$^{c}$ &  supported in part by the MINERVA Gesellschaft f\"ur Forschung GmbH, the Israel Science
          Foundation (grant no. 293/02-11.2) and the U.S.-Israel Binational Science Foundation \\
$^{d}$ &  supported by the German-Israeli Foundation and the Israel Science Foundation\\
$^{e}$ &  supported by the Italian National Institute for Nuclear Physics (INFN) \\
$^{f}$ &  supported by the Japanese Ministry of Education, Culture, Sports, Science and Technology
          (MEXT) and its grants for Scientific Research\\
$^{g}$ &  supported by the Korean Ministry of Education and Korea Science and Engineering
          Foundation\\
$^{h}$ &  supported by the Netherlands Foundation for Research on Matter (FOM)\\
$^{i}$ &  supported by the Polish State Committee for Scientific Research, grant no.
          620/E-77/SPB/DESY/P-03/DZ 117/2003-2005 and grant no. 1P03B07427/2004-2006\\
$^{j}$ &  partially supported by the German Federal Ministry for Education and Research (BMBF)\\
$^{k}$ &  supported by RF Presidential grant N 8122.2006.2 for the leading
          scientific schools and by the Russian Ministry of Education and Science through its grant
          Research on High Energy Physics\\
$^{l}$ &  supported by the Spanish Ministry of Education and Science through funds provided by
          CICYT\\
$^{m}$ &  supported by the Particle Physics and Astronomy Research Council, UK\\
$^{n}$ &  supported by the US Department of Energy\\
$^{o}$ &  supported by the US National Science Foundation. Any opinion,
findings and conclusions or recommendations expressed in this material
are those of the authors and do not necessarily reflect the views of the
National Science Foundation.\\
$^{p}$ &  supported by the Polish Ministry of Science and Higher Education
as a scientific project (2006-2008)\\
$^{q}$ &  supported by FNRS and its associated funds (IISN and FRIA) and by an Inter-University
          Attraction Poles Programme subsidised by the Belgian Federal Science Policy Office\\
$^{r}$ &  supported by the Malaysian Ministry of Science, Technology and
Innovation/Akademi Sains Malaysia grant SAGA 66-02-03-0048\\
\end{tabular}
\newpage

\pagenumbering{arabic}
\pagestyle{plain}
\section{Introduction}
\label{sec-int}

Heavy-quark production
in $ep$ interactions in the deep inelastic scattering (DIS) regime is
dominated by the interaction  between the exchanged virtual photon
and a gluon within the proton, the so-called Boson Gluon Fusion
(BGF) mechanism.
Heavy-quark production provides a twofold test of
perturbative quantum chromodinamics (pQCD): a study of
the BGF process and the higher order corrections to it,
and an independent check of the validity of the gluon density in the
proton extracted from the inclusive DIS data.
Of the two heavy quarks whose production is accessible by HERA,
$c$ and $b$, the latter is strongly suppressed due to its smaller
electric charge and  larger mass. This paper reports a study of
$c$-quark production.

A charm quark in the final state is identified by the presence of a
corresponding charmed hadron.
This paper studies the production of the pseudo-scalar mesons
$D^0$, $D^+$, $D_s^+$ and the vector meson $D^{*+}$
from the decays
$D^0\rightarrow K^-\pi^+$, $D^+\rightarrow K^-\pi^+\pi^+$,
$D_s^+\rightarrow  \phi\pi^+ \rightarrow K^+K^-\pi^+$ and
$D^{*+}\rightarrow D^0\pi^+\rightarrow K^-\pi^+\pi^+$
(the charge conjugated modes are implied throughout this paper).
Since a $D$ hadron is measured and not the $c$ quark itself, any comparison
with pQCD requires a modelling of the $c \rightarrow D$
fragmentation.
A consequence of the QCD factorisation theorem \cite{factheo} is that
the ``hard'' (pQCD governed) $c$-production mechanism is independent of the
``soft'' fragmentation process.
Measurements of $D$-hadron cross sections provide therefore information about both $c$-quark
production and its fragmentation.

This paper presents a complete study of $D$-meson production in DIS at HERA:
measurements of $c$-quark fragmentation ratios and
fractions with unprecedent precision, $D$-meson differential cross sections and the
charm contribution, $F_2^{c\bar{c}}$,
to the proton structure function $F_2$.
It addresses the universality of fragmentation and tests the predictions of
pQCD for charm production.
The data sample used was taken by the ZEUS detector during the years
1998 -- 2000.
The fragmentation measurements follow closely those reported recently by ZEUS
in the photoproduction regime \cite{epj:c44:351}.
Using a variety of $D$ mesons, the pQCD analysis complements the
study done with $D^{*+}$  in the same data sample
\cite{pr:d69:012004}.
Measurements of $D^*$ cross sections are only used in this paper for the
extraction of the fragmentation parameters.

Similar measurements of the properties of $c$-quark fragmentation in DIS
have also been performed by the H1 collaboration \cite{epj:c38:447}.
Other previous measurements of charm production in DIS with pQCD analyses
used the $D^{*+}$ meson
\cite{pr:d69:012004,epj:c12:35,pl:b407:402,pl:b528:199,zfp:c72:593,np:b545:21}
or inclusive lifetime tags \cite{epj:c40:349}. There are also several
measurements of charm photoproduction \cite{pl:b481:213,epj:c6:67,pl:b401:192,
pl:b621:56,np:b545:21,np:b472:32}.

\section{Experimental set-up}
\label{sec-exp}

The analysis was performed with data taken from 1998 to 2000, when HERA
collided electrons or positrons with energy $E_e =$ 27.5~GeV on protons of
energy $E_p =$ 920~GeV. The results are based on $e^-p$ and $e^+p$ samples
corresponding to integrated luminosities of $16.7\pm 0.3$ pb$^{-1}$ and
$65.1\pm 1.5$ pb$^{-1}$, respectively.\footnote{Hereafter, both electrons and
positrons are referred to as electrons, unless explicitly stated otherwise.}

\Zdetdesc

\Zctddesc\ZcoosysfnB

\Zcaldesc

The position of the scattered electron was determined by combining information
from the CAL, the small-angle rear tracking detector (SRTD)~\cite{nim:a401:63}
and the hadron-electron separator (HES)~\cite{nim:a277:176}.

The luminosity was measured from the rate of the bremsstrahlung process
$ep~\rightarrow~e\gamma p$, where the photon was measured in a
lead--scintillator calorimeter~\cite{desy-92-066,*zfp:c63:391,*acpp:b32:2025}
placed in the HERA tunnel at $Z=-107~{\rm m}$.

\section{Theoretical predictions}
\label{sec:theory}

The next-to-leading order (NLO) QCD predictions for $c\bar{c}$ cross sections were
obtained using the HVQDIS program~\cite{pr:d57:2806} based on the so-called
fixed-flavour-number scheme (FFNS). In this scheme, only light partons
($u, d, s, g$) are included in the proton parton density functions (PDFs) which
obey the DGLAP equations
\cite{sovjnp:15:438,*sovjnp:20:94,*np:b126:298,*jetp:46:641},
and the $c\bar{c}$ pair is produced via the BGF
mechanism~\cite{np:b452:109,*pl:b353:535} with NLO
corrections~\cite{np:b392:162,*np:b392:229}.
This calculation is expected to be valid~\cite{epj:c18:547}
 in the kinematic range of this
measurement, $1.5 < Q^2 < 1000$ GeV$^2$, where $Q^2$ is the
negative of the four-momentum transfer squared, hereafter referred to as photon
virtuality.

The following inputs have been used to obtain the predictions for $D$-meson
production at NLO using the program HVQDIS. The FFNS variant of the ZEUS-S
NLO QCD fit~\cite{pr:d67:012007,*misc:www:zeus2002} to structure-function
data was used as the parameterisation of the proton PDFs. In this fit
the three-flavour QCD scale $\Lambda_{\rm QCD}$ was set to
$\Lambda^{(3)}_{\rm QCD}=0.363$~GeV and the mass of the charm
quark was set to 1.35~GeV; the same mass and $\Lambda^{(3)}_{\rm QCD}$
were therefore used in the HVQDIS calculation. The renormalisation and
factorisation scales were set to $\mu = \sqrt{Q^2+4m_c^2}$ for charm
production both in the fit and in the HVQDIS calculation. The charm
fragmentation to the particular $D$ meson was carried out using the Peterson
function~\cite{pr:d27:105}. The values used for the hadronisation fractions
to $D$ mesons, $f(c \to D)$, were those measured in this paper, and the
Peterson parameter, $\epsilon$, was set to 0.035\cite{np:b565:245}.
The effect of $J/\psi$ production was found to be
negligible~\cite{epj:c25:41,epj:c6:603}.

\section{Kinematic reconstruction and event selection}
\label{sub:ds}

The kinematic variables $Q^2$, the Bjorken scaling variable, $x$
(in the quark-parton model $x$ can be interpreted as the fraction
 of proton momentum carried by the struck quark),
and the fraction of the electron energy
transferred to the proton in the rest frame of the proton,
 $y$, can be reconstructed using a
variety of methods, whose accuracy depends on the variable of interest and its
range:

\begin{itemize}

\item
for the electron method (specified with the subscript $e$), the measured energy
and angle of the scattered electron are used;

\item
the double angle (DA) method~\cite{proc:hera:1991:23,*hoeger} relies on the angles of
the scattered electron and of the hadronic system;

\item
the Jacquet-Blondel (JB) method~\cite{proc:epfacility:1979:391} is based
entirely on measurements of the hadronic system;

\item
the $\Sigma$-method~\cite{nim:a361:197} uses both the scattered-electron energy
and angle, and measurements of the hadronic system.

\end{itemize}

The reconstruction of $Q^2$ and $x$ was performed using the $\Sigma$-method,
since it has better resolution at low $Q^2$ than the DA method. At high $Q^2$,
the $\Sigma$-method and the DA method are similar, and both have better
resolution than the electron method. The DA method was used as a systematic
check.

A three-level trigger system was used to select events
online~\cite{zeus:1993:bluebook,proc:chep:1992:222}. At the third level, events
having at least a reconstructed $D^{*+},D^0,D^+$ (only $e^+p$ sample),
$D_s^+$ or $\Lambda_c^+$ candidate,
as well as a scattered-electron candidate, were kept for further analysis.
The efficiency of the online reconstruction for any of the above hadrons,
determined relative to an inclusive DIS trigger, was generally above $95\%$.

The events were selected offline~\cite{epj:c21:443,epj:c12:35} using
the following cuts:

\begin{itemize}

\item
the scattered electron was identified using a neural-network
procedure~\cite{nim:a365:508,nim:a391:360}. Its energy, $E_{e^{'}}$, was
required to be larger than 10~GeV;

\item
$y_{e}\> \leq\> 0.95$ (where $y_e$ is $y$ reconstructed with the electron
method) and $y_{\mathrm{JB}}\> \geq\>  0.02$ 
(where $y_{\mathrm{JB}}$ is $y$ reconstructed with the JB method).
The former condition removes events where fake electrons are 
found in the FCAL and the latter rejects events where the 
hadronic system cannot be measured precisely, in order to 
reconstruct the kinematic variables;

\item
$40$ $\> \leq\>$ $\delta$ $\> \leq\>$ $65$ GeV, where
$\delta=\sum E_i(1-\cos\theta_i)$
and   $E_i$ and $\theta_i$ are the energy and the polar angle of the $i^{th}$
energy-flow object (EFO)~\cite{briskinu:phd:1998} reconstructed from charged
tracks, as measured in the CTD, and energy clusters measured in the CAL.
The sum $i$ runs over all EFOs;

\item
a primary vertex position determined from the tracks fitted to the vertex in
the range $|Z_{\rm vertex}| < 50$ cm;

\item
the impact point ($X$, $Y$) of the scattered electron on the RCAL was required to lie
outside the region \mbox{26 $\times$ 14 cm$^2$} centred on $X=Y=0$.

\end{itemize}

The angle of the scattered  electron was determined using either its impact
position on the CAL inner face or a reconstructed track in the CTD.
When available, SRTD and HES were also used.
The energy of the scattered electron was corrected for non-uniformity
effects caused by cell and module boundaries.

The selected kinematic region was \mbox{$1.5 < Q^2 < 1000$ GeV$^2$} and
\mbox{$0.02 < y < 0.7$}.


\section{Reconstruction of charm hadrons}
\label{sec-rec}

The production of $\dsp$, $\dz$, $\dc$, and $\dssp$ charm mesons was measured
in the range of transverse momentum $p_T(D)>3\gev$ and pseudorapidity
$|\eta(D)|<1.6$.
For the $\dssp$, the $p_T(\dssp)$ requirement was relaxed to $p_T(\dssp)>2\gev$,
as the constraint provided by the $K^+K^-$ pair coming from a $\phi$ meson kept
the combinatorial background at acceptable levels.
The reconstruction of the $\lcp$ baryon was attempted using the decay
$\lcp\rightarrow K^- p \pi^+$. The signal achieved had a statistical
significance of around three standard deviations, and therefore it was not used.

The charm mesons were reconstructed using tracks measured
in the CTD and assigned to the reconstructed event vertex.
To ensure good momentum resolution, each track was required
to reach at least the third superlayer of the CTD.
Further background reduction was achieved by imposing cuts on the transverse
momenta and decay angles of the charm-hadron decay products. The cut values
were optimised using Monte Carlo (MC) simulation to enhance signal 
over background ratios while keeping acceptances high.

The cross sections of the $D^0$ and $D^{*+}$ mesons and related quantities
involved in the measurements of fragmentation properties can be obtained from
the combination of three independent samples~\cite{epj:c44:351}: those of
$D^0$ candidates with and without a ``$\Delta M$'' tag and that of
``additional'' $D^{*+}$ candidates. The samples are described below. The
rationale for this division~\cite{epj:c44:351} will become apparent in
Sections~\ref{sec-ratio} and~\ref{sec-ff}.
%
%
\subsection{Reconstruction of $D^0$ mesons}
\label{subsec-d0} 
The $D^{0}$ mesons were reconstructed from the decay channel
$D^{0}\rightarrow K^{-}\pi^{+}$. In each event, $D^0$ candidates were formed
from pairs of tracks with opposite charges and $p_T > 0.8~$GeV. The nominal
kaon and pion masses were assumed in turn for each track. To reduce the
combinatorial background, a further cut was applied in the angle,
$\theta^*(K)$, between the kaon in the $D$-candidate rest frame and the
$D$-candidate line of flight in the laboratory frame, $|\cos~\theta^*(K)|<0.85$.
The $D^0$ candidates were separated into two groups. The $\Delta M$ tag group
consists of $D^0$ candidates that, when combined with a third track that could be
a ``soft'' pion $(\pi_s)$ in a $D^{*+} \rightarrow D^0\pi_s^+$ decay,
have $\Delta M = M(K\pi\pi_s)-M(K\pi)$ in the range
$0.143< \Delta M < 0.148~$GeV. The soft pion was required to have
$p_T > 0.2$~GeV and charge opposite to that of the kaon. This cut was raised to
$p_T > 0.25$ GeV for a data subsample, corresponding to an integrated luminosity
of $17$~pb$^{-1}$, for which the track reconstruction efficiency at low momentum
was smaller due to the operating conditions of the CTD \cite{Bailey:2001yn}.
For the untagged $D^0$ candidates, the incorrect assignment of the
pion and kaon masses to the two tracks produces a wider reflected signal. This
reflection was estimated from the $D^0$ candidates with a $\Delta M$ tag
and normalised to the ratio of numbers of $D^0$ without and with $\Delta M$ tag;
it was then subtracted from the untagged $D^0$ candidates.

\begin{figure}[hbtp]
\epsfysize=18cm
\vspace*{-1.0cm}
\centerline{\mbox{\epsfig{file=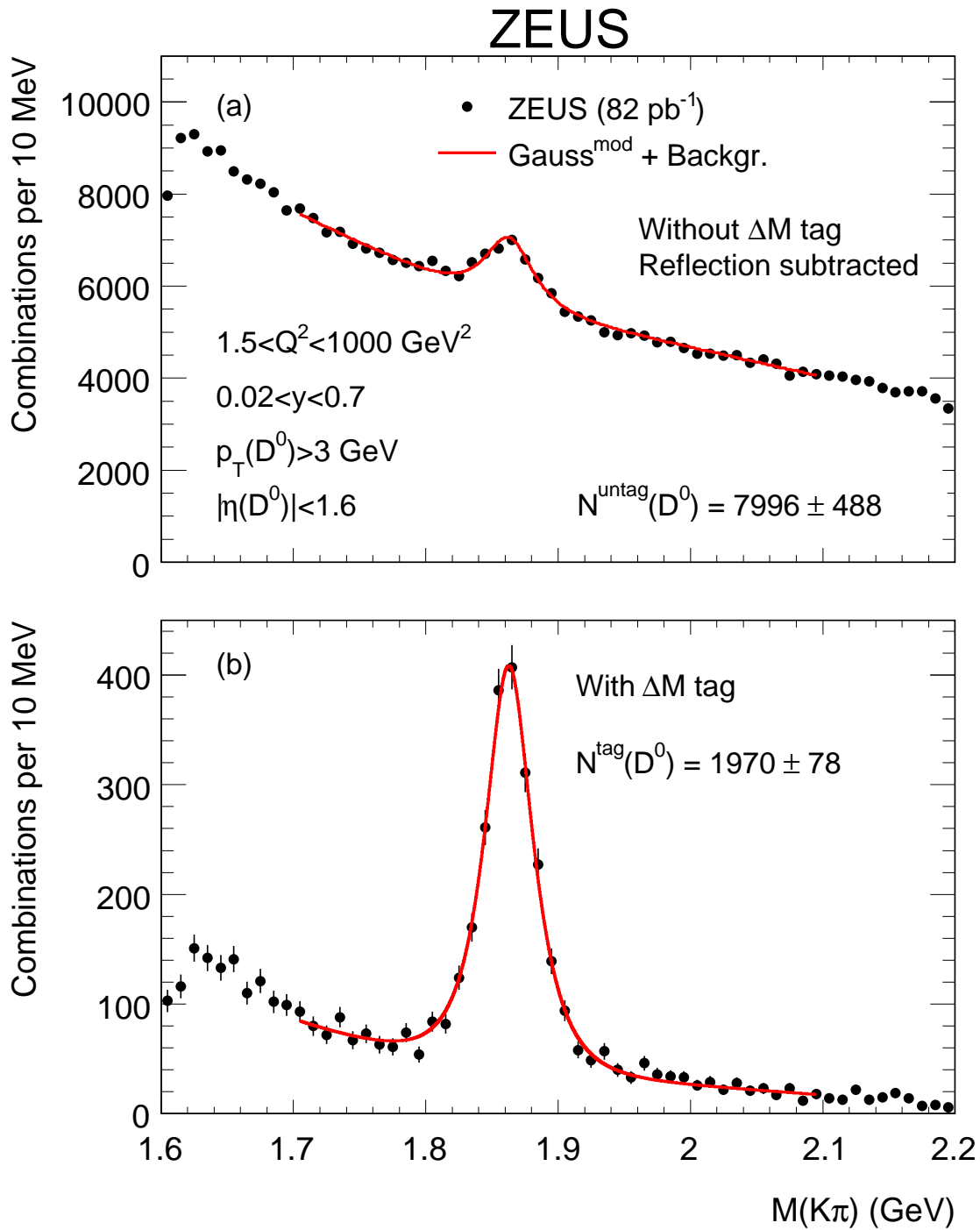,height=20cm}}}
\caption{
The $M(K^- \pi^+)$ distributions (dots) for (a) the $D^0$ candidates
without $\Delta M$ tag, obtained after reflection subtraction,
and for (b) the $D^0$  candidates
with $\Delta M$ tag.
The lowest and highest mass bins are affected by the trigger selection.
The solid curves represent a fit to the sum of a modified Gaussian
function and a background function.
}
\label{fig:d0}
\end{figure}

Figure \ref{fig:d0} shows the $M(K\pi)$ distributions for untagged $D^0$
candidates after the reflection subtraction and for tagged $D^0$ candidates.
The distributions were fitted simultaneously assuming that both have the same
shape and are described by the ``modified'' Gaussian function \cite{epj:c44:351}:

\begin{equation}
{\rm Gauss}^{\rm mod}(M_0,\sigma)\propto e^{[-0.5 \cdot x^{1+1/(1+0.5 \cdot x)}]},
\label{eq:gausmod}
\end{equation}

where $x=|[M(K^-\pi^+)-M_0]/\sigma|$. The background shape in the fit was
described \cite{epj:c44:351} by the form $[A+B\cdot M(K^- \pi^+)]$ for
$M(K^- \pi^+)>1.86\gev$ and
$[A+B\cdot M(K^- \pi^+)]\cdot \exp\{C\cdot[M(K^- \pi^+)-1.86]\}$ for
$M(K^- \pi^+)<1.86\gev$. The free parameters $A$, $B$ and $C$ were assumed to be
independent for the two $M(K^- \pi^+)$ distributions. Clear signals are seen at
the nominal value of $M(D^0)$~\cite{jp:g33:1}.
The number of untagged (tagged) $D^0$ mesons
yielded by the fit was $N^{\rm untag}(D^0) = 7996 \pm 488$
($N^{\rm tag}(D^0) = 1970 \pm 78$).
%
%
\subsection{Reconstruction of ``additional'' $D^{*+}$ mesons}
\label{subsec-ds1}
The $\dsp \rightarrow \dz \pi^+_s$ events with $p_T(\dsp)>3\gev$ and
$|\eta(\dsp)|<1.6$ can be considered as a sum of two subsamples: events with
the $\dz$ having $p_T(\dz)>3\gev$ and $|\eta(\dz)|<1.6$, and events with the
$\dz$ outside of that kinematic range. The former sample is a subset of $\dz$
mesons reconstructed with $\Delta M$ tag, as discussed above. The latter
sample of additional $\dsp$ mesons was obtained using the same
$D^0 \rightarrow K^- \pi^+$ decay channel. In each event, pairs of tracks with
$p_T>0.4~$GeV were combined to form a $D^0$ candidate. Only combinations with
invariant mass $1.80< M(K\pi)< 1.92~$GeV were considered. The $D^0$ candidates
were required to have $p_T(D^0)<3.0~$GeV or $|\eta(D^0)|>1.6$.
A third track with $p_T>0.2~$GeV, with charge opposite to that of the kaon
track and assumed to have the pion mass, was combined with the $D^0$ candidate
to form an additional $D^{*\pm}$ candidate. Here again the cut value was
$p_T > 0.25$ GeV for the data subsample for which the track reconstruction
efficiency at low momentum was smaller.

\begin{figure}[hbtp]
\epsfysize=18cm
\vspace*{-2.0cm}
\centerline{\mbox{\epsfig{file=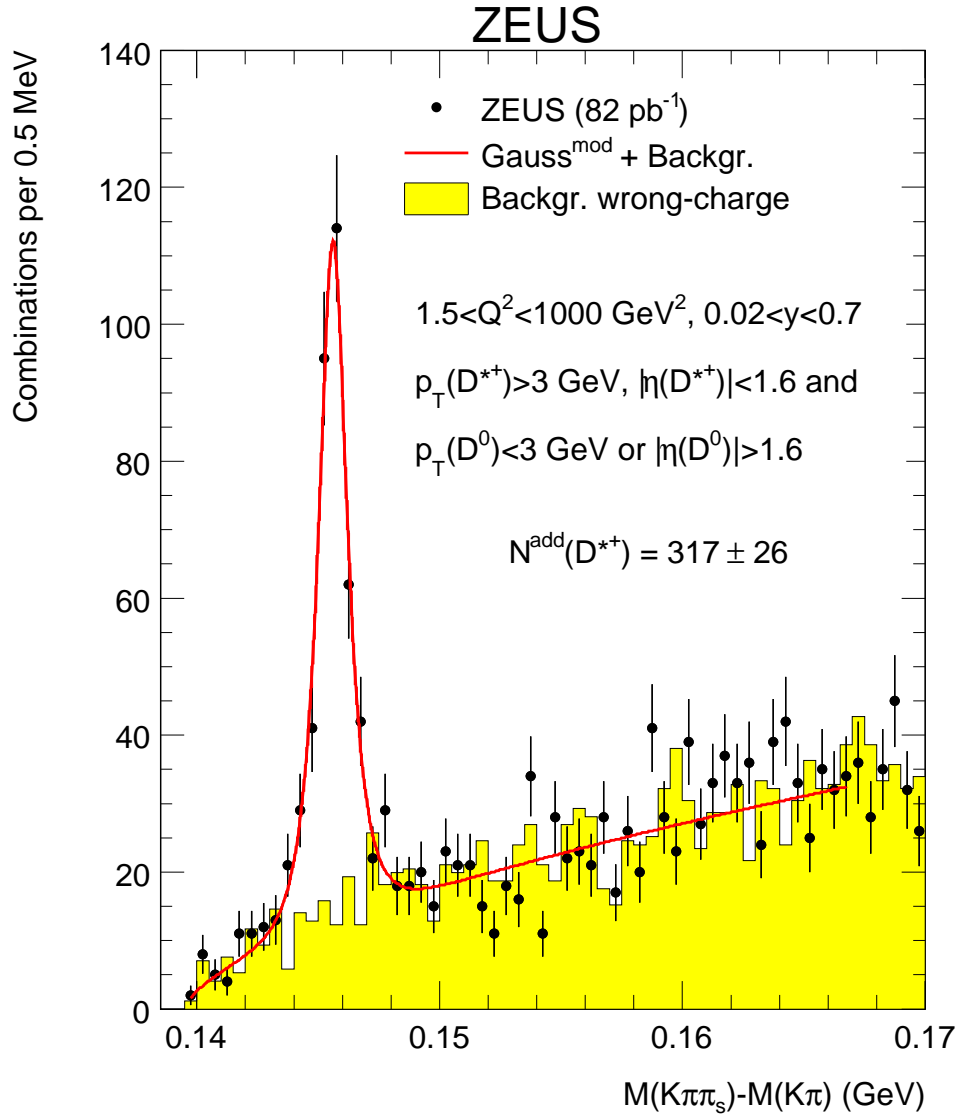,height=17cm}}}
\caption{
The distribution of the mass difference,
$\Delta M=M(K^- \pi^+ \pi_s^+)-M(K^- \pi^+)$, for the ``additional''
$D^{*+}$  candidates (dots).
The histogram
shows the $\Delta M$ distribution for wrong-charge combinations.
For illustration, the solid curve represents a fit to the sum of a
modified Gaussian function and a background function.
}
\label{fig:ds}
\end{figure}

Figure~\ref{fig:ds} shows the distribution of the mass difference
$\Delta M = M(K\pi\pi_s)-M(K\pi)$ for additional $D^*$ candidates. A clear signal is seen
around the nominal value of $M(D^{*\pm})-M(D^0)$~\cite{jp:g33:1}. The
combinatorial background under the signal was estimated from the
mass-difference distribution of the wrong-charge combinations, in which both
tracks associated to the $D^0$ candidate have the same charge and the third
track has opposite charge. The number of reconstructed additional $D^{*+}$
mesons was calculated by subtracting the wrong-charge $\Delta M$ distribution,
after normalising it to the number of right-charge candidates in the region
$0.150< \Delta M < 0.170~$GeV. The subtraction was done in the signal region
$0.143< \Delta M < 0.148~$GeV and yielded $N^{\rm add}(D^{*+})=317\pm 26$.
%
%
\subsection{Reconstruction of $D^+$ mesons} 
\label{subsec-dch}
The $D^+$ mesons were reconstructed from the decay channel
$D^+\rightarrow K^{-}\pi^{+}\pi^{+}$.
 The analysis for this meson was done
using the $e^+p$ data sample only, where the $D^+$ third level 
trigger  logic was implemented. In each event, 
two tracks with the same charge and a third track
with opposite charge were combined to form a $D^+$ candidate. The tracks with
the same charge were assigned the pion mass and required to have
$p_T(\pi)>0.5$ GeV. For the remaining track, the kaon mass was assumed and
$p_T(K)>0.7$ GeV was required. Combinatorial background was further suppressed
by requiring $\cos\theta^*(K)>-0.75$. Background from $D^{*+}$ and
$D_s^+\rightarrow \phi\pi^+$ with $\phi\rightarrow  K^{+}K^{-}$ was suppressed
by applying suitable cuts~\cite{epj:c44:351}.

\begin{figure}[hbtp]
\epsfysize=18cm
\vspace*{-2.0cm}
\centerline{\mbox{\epsfig{file=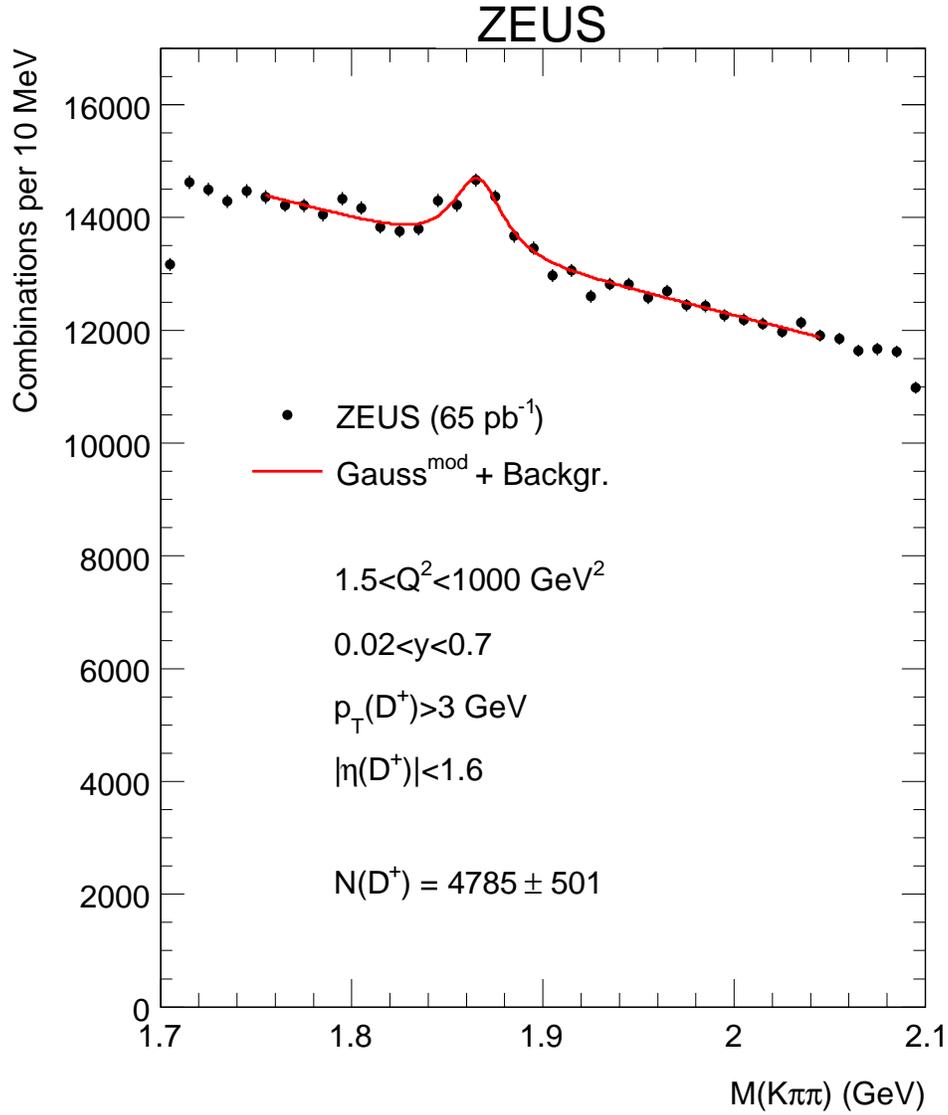,height=17cm}}}
\caption{
The $M(K^-\pi^+\pi^+)$ distribution for
the $D^+$ candidates (dots).
The lowest and highest mass bins are affected by the trigger selection.
The solid curve represents a fit to the sum of a modified Gaussian
function and a linear background function.
}
\label{fig:dc}
\end{figure}

Figure~\ref{fig:dc} shows the $M(K^-\pi^+\pi^+)$ distribution for the $\dc$
candidates. A clear signal is seen at the nominal value of $D^+$ mass
\cite{jp:g33:1}. The mass distribution was fitted to a sum of the modified
Gaussian function (Eq.~(\ref{eq:gausmod})) describing the signal and a linear
function describing the non-resonant background. The number of reconstructed
$\dc$ mesons yielded by the fit was $N(\dc)=4785\pm501$.
%
%
\subsection{Reconstruction of $D_s^+$ mesons}
\label{subsec-dss}
The $D_s^+$ mesons were reconstructed from the decay channel
$D_s^{+}\rightarrow \phi \pi^{+}$  with $\phi\rightarrow K^{+}K^{-}$. In each
event, two tracks with opposite charges were each assigned the kaon mass and
combined to form the $\phi$ candidates if the invariant mass $M(KK)$ was
within $\pm 8$~MeV of the nominal $\phi$ mass~\cite{epj:c44:351}. A third
track, assumed to be a pion, was combined with the $\phi$ candidate, yielding
the $D_s^+$ candidate. Only tracks with $p_T(\pi)>0.5$ GeV and $p_T(K)>0.7$ GeV
were considered. To reduce the combinatorial background further, a cut was
applied in the angle, $\theta^*(\pi)$, between the pion in the $KK\pi$ rest
frame and the $KK\pi$ line of flight in the laboratory frame,
$\cos \theta^*(\pi)<0.85$, and in the angle, $\theta'(K)$, between one of the
kaons and the pion in the $KK$ rest frame, $|\cos^3\theta'(K)|>0.1$.

\begin{figure}[hbtp]
\epsfysize=18cm
\vspace*{-2.0cm}
\centerline{\mbox{\epsfig{file=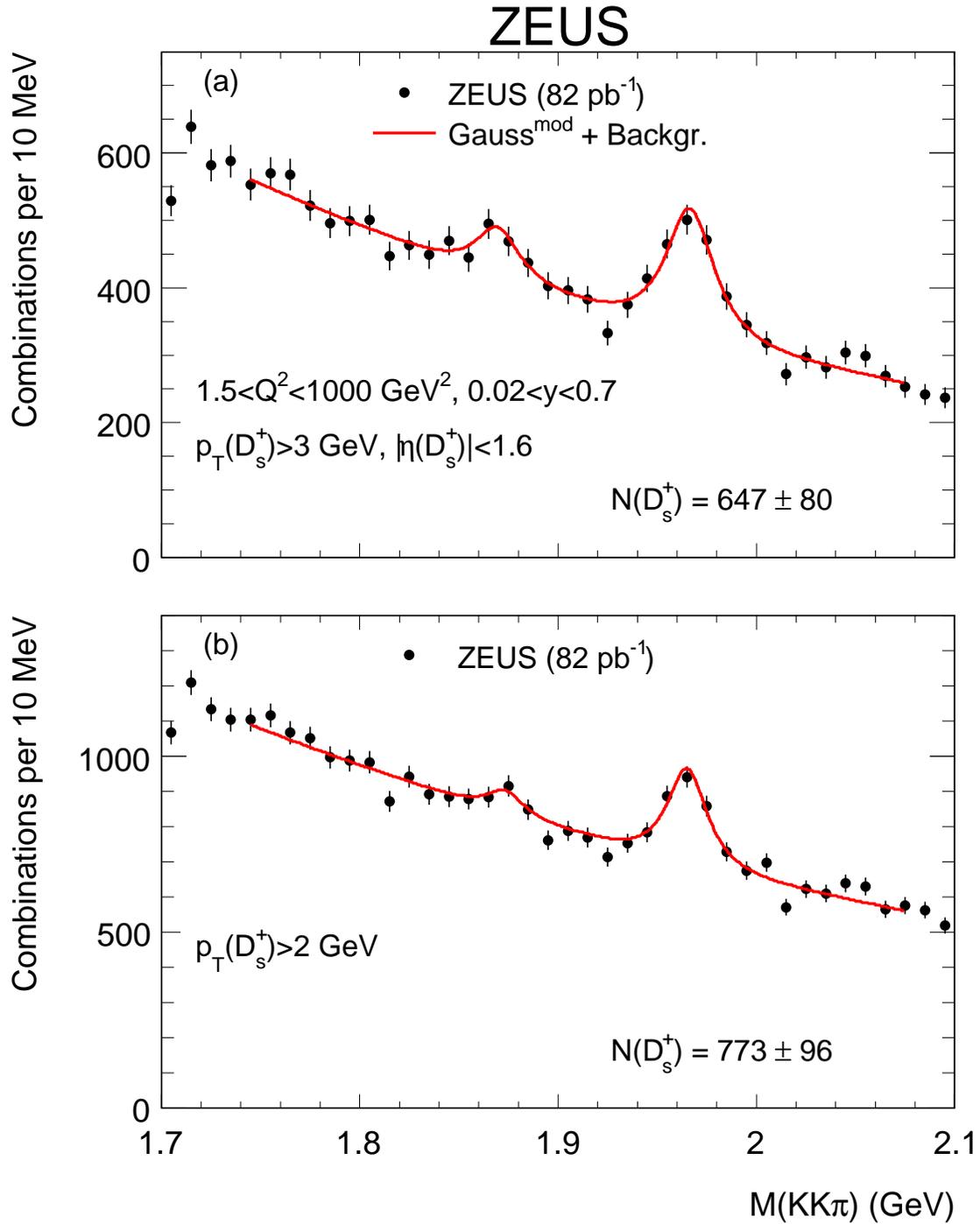,height=20cm}}}
\caption{
The $M(K^+ K^-\pi^+)$ distribution for
(a) $D_s^+$ candidates with $p_T(D_s^+)>3$ GeV and
(b) $D_s^+$ candidates with $p_T(D_s^+)>2$ GeV.
The lowest mass bins are affected by the trigger selection.
The solid curves represent  fits to the sum of
two modified Gaussian functions and an exponential
background function. The first peak in both distributions
is from $D^+$ decaying through the same channel.
}
\label{fig:dss}
\end{figure}

Figure~\ref{fig:dss} shows the $M(K^+ K^- \pi^+)$ distributions for the $\dssp$
candidates after all cuts.
 A clear signal is seen at the nominal $\dssp$ mass~\cite{jp:g33:1}.
There is also a smaller signal around the nominal $\dc$ mass as expected from
the decay $\dc \rightarrow \phi\pi^+$. The mass distribution was fitted to a
sum of two modified Gaussian functions (Eq.~(\ref{eq:gausmod})) describing the
signals and an exponential function describing the non-resonant background.
The number of reconstructed $\dssp$ mesons yielded by the fit was
$N(\dssp)=647\pm 80$, for $p_T(\dssp)>3$ GeV and $N(\dssp)=773\pm 96$, for 
$p_T(\dssp)>2$ GeV.

\section{Acceptance corrections}
\label{sec:evsim} 

The acceptances were calculated using the {\sc Rapgap 2.08}~\cite{cpc:86:147}
MC model and checked with {\sc Herwig 6.3}~\cite{hep-ph-9912396,*cpc:67:465}. The
{\sc Rapgap} MC model was interfaced with {\sc Heracles 4.6.1}~\cite{cpc:69:155}
in order to incorporate first-order electroweak corrections. The generated
events were then passed through a full simulation of the detector, using
{\sc Geant 3.13}~\cite{tech:cern-dd-ee-84-1}, processed and selected with
the same programs as used for the data.

The MC models were used to produce charm by the BGF and the resolved photon
processes. In the latter the virtual photon behaves as a hadron-like 
source of partons, one of which interacts with a parton of the initial proton.
 The CTEQ5L~\cite{epj:c12:375} and GRV-LO~\cite{pr:d46:1973} PDFs were
used for the proton and the photon, respectively. The charm-quark mass was set
to $1.5$~GeV. Both the {\sc Rapgap} and {\sc Herwig} MCs use LO matrix elements
with leading-logarithmic parton showers. Charm fragmentation is implemented
using either the Lund string fragmentation (in {\sc Rapgap}) or a cluster
fragmentation~\cite{np:b238:492} model (in {\sc Herwig}). It was checked that
both MC samples, {\sc Rapgap} and {\sc Herwig}, give a reasonable description of
the data for  DIS and $D$-meson variables when compared at detector-level.

For a given observable $Y$, the production differential
cross sections were determined using

\begin{equation}
\frac {d\sigma}{dY} =
\frac {N(D) } {\mathcal {A} \cdot \mathcal {L} \cdot \mathcal {B} \cdot
\Delta Y},
\nonumber
\end{equation}

where $N(D)$ is the number of reconstructed $D$ mesons in a bin of size
$\Delta Y$. The reconstruction acceptance $\mathcal {A}$ takes into
account small admixtures in the reconstructed signals from other decay
modes, migrations, efficiencies and QED radiative effects for that bin,
$\mathcal {L}$ is the integrated luminosity and $\mathcal {B}$ is the
branching ratio~\cite{jp:g33:1} for the decay channel used in the
reconstruction (see Table~\ref{tab:branching}).
The total production cross sections were determined using

\begin{equation}
\sigma =
\frac {N(D) } {\mathcal {A} \cdot \mathcal {L} \cdot \mathcal {B}},
\nonumber
\end{equation}

where $N(D)$ and $\mathcal {A}$ are now for the whole kinematical
range of the measurement.

The reconstruction acceptances were calculated with {\sc Rapgap} and vary
depending on the particle and the kinematic region of the measurement. For
$1.5<Q^2<1000$ GeV$^2$, $0.02<y<0.7$, transverse momenta $p_T(\dz,\dc)>3\gev$,
$p_T(\dssp)>2\gev$ and pseudorapidity $|\eta(D)|<1.6$ the overall
acceptances were $\approx 42\%$, $\approx 26\%$ and $\approx 17\%$ for
$\dz$, $\dc$, and $\dssp$ mesons, respectively.

The relative $b$-quark contributions, predicted by the MC simulation using
branching ratios of $b$-quark decays to the charmed hadrons measured at
LEP~\cite{pl:b388:648,epj:c1:439}, were subtracted from all measured cross
sections. The subtraction of the $b$-quark contribution reduced the measured
cross sections by 3.1\% for the $\dz$ and $D^+$ and
 4.3\% for the $D_s^{+}$ and changed the measured charm fragmentation
ratios and fractions by less than $1\%$.

\section{Charm-meson production cross sections}
\label{sec-xsec}

Charm-meson cross sections were calculated using the reconstructed signals for
the process $e p\rightarrow e D X$ in the kinematic region $1.5<Q^2<1000\gev^2$,
$0.02<y<0.7$, $p_T(D)>3\gev$ (for the $D_s^+$ also $p_T(D_s^+)>2\gev$) and
$|\eta(D)|<1.6$.

The systematic uncertainties presented in this and the following sections
will be discussed in Section~\ref{sec-syst}. The third set of uncertainties
quoted for the measured cross sections and charm fragmentation ratios and
fractions are due to the propagation of the relevant branching-ratio
uncertainties.

The following cross sections were measured:

\begin{itemize}
\item{ 
the production cross section for $\dz$ mesons not originating from the
$\dsp \rightarrow \dz \pi^+_s$ decays, hereafter called untagged $D^0$ mesons, is:

$$\sigma^{\rm untag}(\dz)=5.56\pm0.35({\rm stat.})^{+0.32}_{-0.26}({\rm syst.})\pm0.10({\rm br.})\,{\rm nb};$$
}
\item{
the production cross section for $\dz$ mesons originating
from the $\dsp \rightarrow \dz \pi^+_s$ decays:

$$\sigma^{\rm tag}(\dz)=1.78\pm0.08({\rm stat.})^{+0.12}_{-0.10}({\rm syst.})\pm0.03({\rm br.})\,{\rm nb};$$
}
\item{the production cross section for all $\dz$ mesons:

$$
\sigma(\dz) = \sigma^{\rm untag}(\dz)+\sigma^{\rm tag}(\dz)=
7.34 \pm 0.36(\mbox{stat.})^{+0.35}_{-0.27}(\mbox{syst.})\pm 0.13 (\mbox{br.}) \mbox{ nb};
$$
}
\item{
the production cross section for additional $\dsp$ mesons:

$$\sigma^{\rm add}(\dsp)=0.518\pm0.046({\rm stat.})^{+0.051}_{-0.046}({\rm syst.})\pm0.01({\rm br.})\,{\rm nb}.$$

The production cross section for $\dsp$ mesons in the kinematic range
$p_T(\dsp)>3\gev$ and $|\eta(\dsp)|<1.6$, $\sigma^{\rm kin}(\dsp)$, 
is given by the sum $\sigma^{\rm add}(\dsp)+\sigma^{\rm tag}(\dz)/\br$:

$$\sigma^{\rm kin}(\dsp)=3.14\pm 0.12({\rm stat.})^{+0.18}_{-0.15}({\rm syst.})\pm0.06({\rm br.})\,{\rm nb};$$
}
\item{
the production cross section for $\dc$ mesons:

$$\sigma(\dc)=2.80\pm0.30({\rm stat.})^{+0.18}_{-0.14}({\rm syst.})\pm0.10({\rm br.})\,{\rm nb};$$
}
\item{ 
the production cross section for $\dssp$ mesons with $p_T(\dssp)>3\gev$:

$$\sigma(\dssp)=1.27\pm0.16({\rm stat.})^{+0.11}_{-0.06}({\rm syst.})^{+0.19}_{-0.15}({\rm br.})\,{\rm nb};
$$
}
\item{
the production cross section for $\dssp$ mesons with $p_T(\dssp)>2\gev$:

$$
\sigma_2(\dssp) =
2.42 \pm 0.30(\mbox{stat.})^{+0.30}_{-0.14}(\mbox{syst.})^{+0.35}_{-0.27}(\mbox{br.}) \mbox{ nb}.
$$
}
\end{itemize}

\section{Charm fragmentation ratios}
\label{sec-ratio}

In this section, the ratio of neutral to charged $D$-meson production rates,
$R_{u/d}$, the strangeness-suppression factor, $\gamma_s$, and the fraction
of charged $D$ mesons produced in a vector state, $P^d_{\rm v}$, are presented
in the kinematic range $1.5<Q^2<1000$ GeV$^2$, $0.02<y<0.7$, $p_T(D)>3\gev$
and $|\eta(D)|<1.6$.

\subsection{Ratio of neutral to charged $D$-meson production rates}
\label{sec-rud}

Neglecting influences from decays of heavier excited $D$ mesons, the ratio of
neutral to charged $D$-meson production rates is given by the ratio of the sum
of $D^{*0}$ and direct $\dz$ production cross sections to the sum of $\dsp$
and direct $\dc$ production cross sections. This ratio can be calculated
as~\cite{epj:c44:351}

$$R_{u/d}=\frac{\sigma^{\rm untag}(D^0)}{\sigma(D^{+})+\sigma^{\rm tag}(D^0)}.$$

Using the measured cross sections, the ratio of neutral to charged $D$-meson
production rates is

$$R_{u/d} =1.22\pm0.11({\rm stat.})^{+0.05}_{-0.02}({\rm syst.})\pm 0.03({\rm br.})\mbox{.}$$

The measured $R_{u/d}$ value agrees with unity, i.e.$\:$it is consistent with
isospin invariance, which implies that $u$ and $d$ quarks are produced equally
in charm fragmentation.

\begin{figure}
\centerline{\epsfig{file=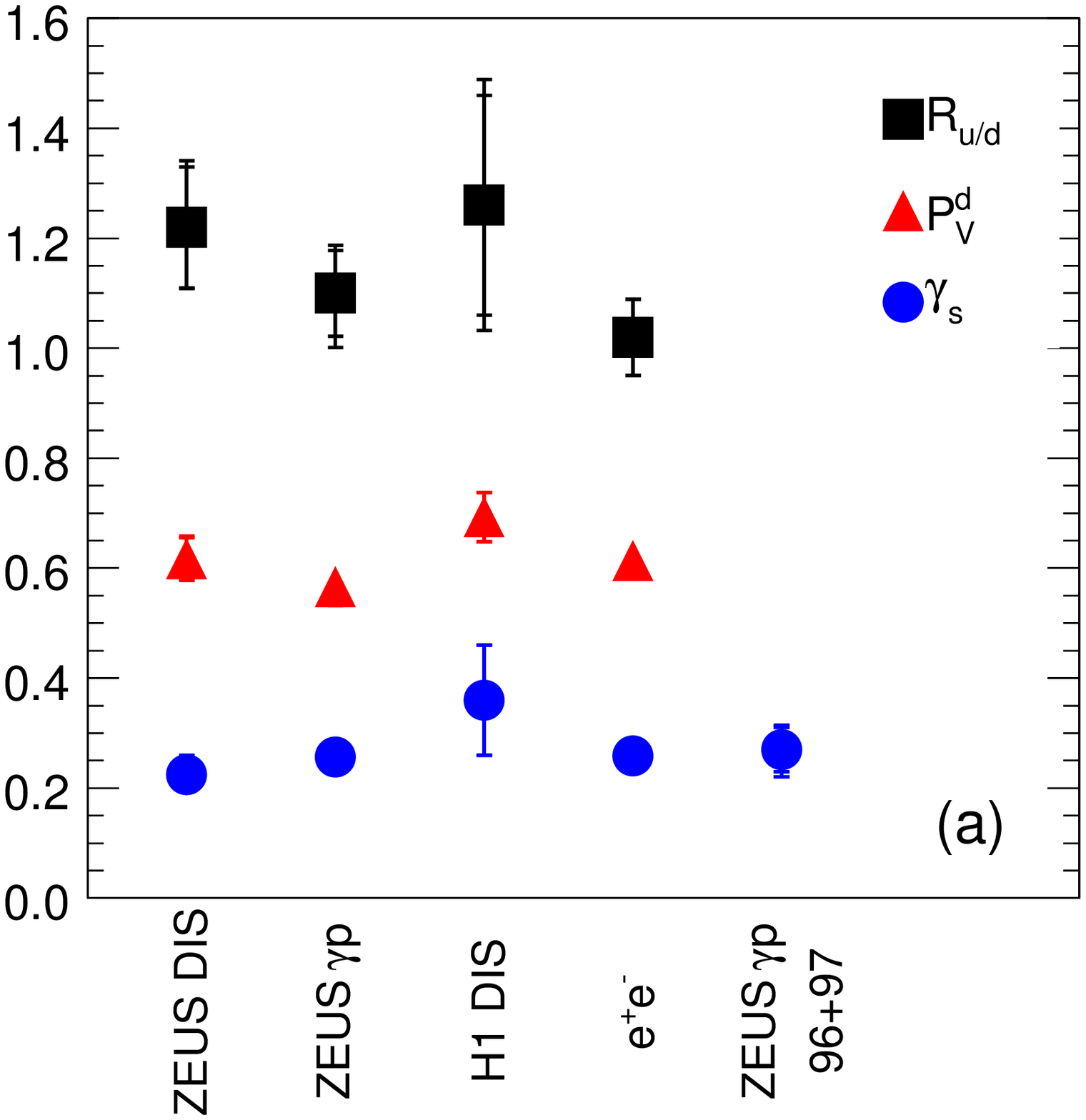,height=9cm}}
\centerline{\epsfig{file=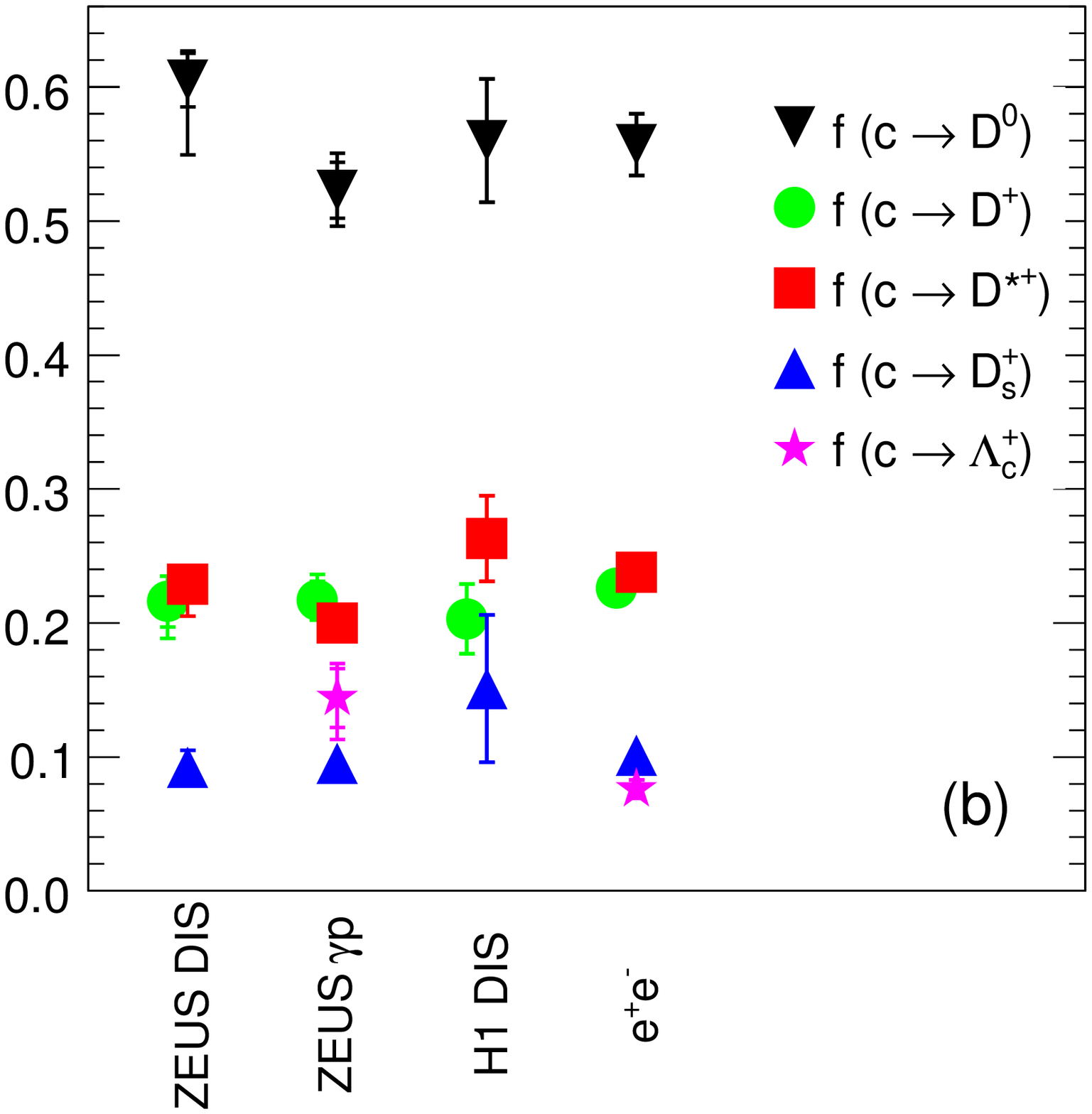,height=9cm}}
\caption[ratios_fractions]
{(a) The ratio of neutral to charged $D$-meson production rates, $R_{u/d}$,
the strangeness-suppression factor in charm fragmentation, $\gamma_s$, and
the fraction of charged $D$ mesons produced in a vector state,
$P^d_{\rm v}$. (b) The fractions of $c$ quarks hadronising as
$D^+$, $D^0$ and $D^+_s$
charm ground-state mesons,  as $D^{*+}$  mesons
and as $\Lambda_c^+$ baryons.
 The inner error bars show the statistical
uncertainties and the outer bars show the statistical and systematic
uncertainties added in quadrature.
The measurements have further  uncertainties
coming from the different branching ratios involved; their magnitudes are shown in
Tables~\ref{tab:rud},~\ref{tab:gs} and~\ref{tab:pv} for $R_{u/d}$,
$\gamma_s$ and
$P^d_{\rm v}$, respectively, and in Table~\ref{tab:ff} for the fractions.
}
\label{ratios_fractions}
\end{figure}

In Table~\ref{tab:rud} and Fig.~\ref{ratios_fractions}, this measurement is
compared with those obtained in DIS by the H1 collaboration~\cite{epj:c38:447},
in photoproduction~\cite{epj:c44:351} and in $e^+e^-$
annihilations~\cite{hep-ex-9912064p}. All measurements agree within experimental
uncertainties.

\subsection{Strangeness-suppression factor}
\label{sec-gs}

The strangeness-suppression factor for charm mesons is given by the ratio of
twice the production rate of charm-strange mesons to the production rate of
non-strange charm mesons. All $\dsp$ and $D^{*0}$ decays produce either a
$\dc$ or a $D^0$ meson, while all $D_s^{*+}$ decays produce a $\dssp$
meson~\cite{jp:g33:1}. Thus, neglecting decays of heavier excited charm-strange
mesons to non-strange charm mesons, the strangeness-suppression factor can be
calculated as the ratio of twice the $\dssp$ cross section to the sum of $\dz$
and $\dc$ cross sections.

For the comparison of the inclusive $D^+$, $D^0$ and $D^{*+}$ cross sections
with each other, the equivalent phase-space treatment ~\cite{epj:c44:351} was
used. The equivalent $D^+$ and $D^0$ cross sections were defined as the sum of
their direct cross sections, i.e. $D^+$ and $D^0$ not coming from $D^{*}$, and
the contribution from $D^{*+}$ and $D^{*0}$ decays~\cite{epj:c44:351}:

\begin{eqnarray*}
\sigma^{\rm eq}(\dc)&=&\sigma(\dc) + \sigma^{\rm add}(\dsp) \cdot (1-\br),\\
\sigma^{\rm eq}(\dz)&=&\sigma^{\rm untag}(\dz) + \sigma^{\rm tag}(\dz) +
\sigma^{\rm add}(\dsp) \cdot \br + \sigma^{\rm add}(D^{*0})
\mbox{.}
\end{eqnarray*}
It was also assumed that the production cross section for ``additional'' $D^{*0}$ mesons,
producing $D^0$ mesons outside of the nominal kinematic range, is
$\sigma^{\rm add}(D^{*0}) = R_{u/d} \sigma^{\rm add}(D^{*+})$.
The strangeness-suppression factor is then given by

$$\gamma_s = \frac{2 \, \sigma(D^+_s)}{\sigma^{\rm eq}(D^+) + \sigma^{\rm eq}(D^0)}
= \frac{2 \, \sigma(D^+_s)}{\sigma(D^+) + \sigma^{\rm untag}(D^0) + \sigma^{\rm tag}(D^0) + \sigma^{\rm add}(D^{*+})\cdot (1+R_{u/d})}
\mbox{.}$$
Using the measured cross sections, the strangeness-suppression factor is

$$ \gamma_s = 0.225 \pm 0.030({\rm stat.})^{+0.018}_{-0.007}({\rm syst.})^{+0.034}_{-0.026}({\rm br.})
\mbox{.}$$
Table~\ref{tab:gs} and Fig.~\ref{ratios_fractions} compare this measurement
with the values measured in photoproduction~\cite{epj:c44:351}, in DIS by
the H1 collaboration~\cite{epj:c38:447} and in $e^+e^-$
annihilations~\cite{hep-ex-9912064p}. All measurements agree within experimental
uncertainties. The large branching-ratio uncertainties are dominated by the
uncertainties of the ${\dssp}\rightarrow\phi\pi^{+}$ branching ratio.

\subsection{Fraction of charged $D$ mesons produced in a vector state}
\label{sec-pv}

Neglecting influences from decays of heavier excited $D$ mesons, the fraction of
charged $D$ mesons produced in a vector state, $P^d_{\rm v}$, is given by the
ratio of vector to (vector+pseudoscalar) charm meson production cross sections.
The following relation holds~\cite{epj:c44:351}:

$$P^d_{\rm v}
= \frac{\sigma^{\rm tag}(D^0)/\br + \sigma^{\rm add}(D^{*+})}
{\sigma(D^+) + \sigma^{\rm tag}(D^0) + \sigma^{\rm add}(D^{*+})}.$$

Using the measured cross sections, the fraction of charged $D$ mesons produced
in a vector state is

$$P^d_{\rm v} = 0.617\pm0.038({\rm stat.})^{+0.017}_{-0.009}({\rm syst.})\pm0.017({\rm br.})
\mbox{.}$$

The measured $P^d_{\rm v}$ value is  smaller than the naive spin-counting
prediction of $0.75$. Table~\ref{tab:pv} and Fig.~\ref{ratios_fractions}
compare this measurement with the values measured in
photoproduction~\cite{epj:c44:351}, in DIS by the H1 
collaboration~\cite{epj:c38:447} and in $e^+e^-$
annihilations~\cite{hep-ex-9912064p}. All the measurements are consistent.

\section{Charm fragmentation fractions} 
\label{sec-ff}

The fraction of $c$ quarks hadronising as a particular charm meson,
$f(c\rightarrow D)$, is given by the ratio of the production cross section for
the meson to the sum of the production cross sections for all charm ground
states that decay weakly, $\sigma_{\rm gs}$. In addition to the measured $\dz$,
$\dc$ and $\dssp$ charm ground states, the production cross sections of the
$\lcp$ baryon and of the charm-strange baryons $\Xi^{+}_c$, $\Xi^{0}_c$
and $\Omega^0_c$ should be included in the sum. The production rates for the
latter are expected to be much lower than that of the $\lcp$ due to strangeness
suppression.

The $\sigma(\Lambda_c^+)$ was estimated using the corresponding fragmentation
fraction measured in $e^+e^-$~\cite{hep-ex-9912064p},
$f(c\rightarrow \Lambda_c^+)_{e^+e^-}$, by the relation below: 

\begin{equation}
 f(c\rightarrow \Lambda_c^+)_{e^+e^-} = \sigma(\Lambda_c^+)/\sigma_{\rm gs}
\label{eq-gs2}
\end{equation}

The uncertainty of this procedure was estimated by using
$f(c\rightarrow \Lambda_c^+)$ obtained in photoproduction~\cite{epj:c44:351},
and considering the uncertainty in
$f(c\rightarrow \Lambda_c^+)_{e^+e^-}$~\cite{hep-ex-9912064p}.

The relative rates for the weakly-decaying charm-strange baryons were estimated
from the non-charm sector following the LEP procedure~\cite{zfp:c72:1}. The
measured $\Xi^{-}/\Lambda$ and $\Omega^{-}/\Lambda$ relative rates are
$(6.65\pm 0.28)\%$ and $(0.42\pm 0.07)\%$, respectively~~\cite{jp:g33:1}.
Assuming equal production of $\Xi^{0}$ and $\Xi^{-}$ states and that a similar
suppression is applicable to the charm baryons, the total rate for the three
charm-strange baryons relative to the $\lcp$ state is expected to be about
$14\%$. Therefore, the estimated $\lcp$ production cross section was scaled by a
factor $1.14$ in the sum of the production cross sections. An error of $\pm0.05$
was assigned to the scale factor when evaluating systematic uncertainties.

Using the equivalent $\dz$ and $\dc$ cross sections~\cite{epj:c44:351},
$\sigma_{\rm gs}$ is given by

$$\sigma_{\rm gs} = \sigma^{\rm eq}(D^+) + \sigma^{\rm eq}(D^0) + \sigma(D_s^+)
+ \sigma(\Lambda_c^+)\cdot 1.14 \mbox{,}$$
which can be expressed as

\begin{equation}
\sigma_{\rm gs} = \sigma(D^+) + \sigma^{\rm untag}(D^0) +
\sigma^{\rm tag}(D^0) +
\sigma^{\rm add}(D^{*+})\cdot (1+R_{u/d}) + \sigma(D_s^+)
+ \sigma(\Lambda_c^+)\cdot 1.14 \mbox{.}
\label{eq-gs1}
\end{equation}

Using the measured cross sections and combining Eqs. (\ref{eq-gs1}) and
(\ref{eq-gs2}) yields

$$\sigma_{\rm gs} =
13.7 \pm 0.6 \,({\rm stat.}) ^{+1.4}_{-0.6} \,({\rm syst.})\pm0.6 \,({\rm br.})\,{\rm nb}
\mbox{.}$$

The fragmentation fractions for the measured charm ground state are given by

\begin{eqnarray*}
f(c\rightarrow D^+)&=&\sigma^{\rm eq}(D^+)/\sigma_{\rm gs}
=[\sigma(D^+)+\sigma^{\rm add}(D^{*+})\cdot (1-\br)]/\sigma_{\rm gs},\\
f(c\rightarrow D^0)&=&\sigma^{\rm eq}(D^0)/\sigma_{\rm gs}\\
&=&[\sigma^{\rm untag}(D^0)+\sigma^{\rm tag}(D^0)+\sigma^{\rm add}(D^{*+})\cdot (R_{u/d}+\br)]/\sigma_{\rm gs},\\
f(c\rightarrow D_s^+)&=&\sigma(D_s^+)/\sigma_{\rm gs}.
\end{eqnarray*}

The fragmentation fraction for the $D^{*+}$ is

$$f(c\rightarrow D^{*+})
=\sigma^{\rm kin}(D^{*+})/\sigma_{\rm gs}
= [\sigma^{\rm tag}(D^0)/\br+\sigma^{\rm add}(D^{*+})]/\sigma_{\rm gs}.$$

The open-charm fragmentation fractions, measured in the kinematic region
$1.5<Q^2<1000 \gev^2$, $0.02 < y< 0.7 $, $p_T(D)>3\gev$ and $|\eta(D)|<1.6$,
are summarised in Table~\ref{tab:ff} and Fig.~\ref{ratios_fractions}.
The results are compared with the values obtained in
photoproduction~\cite{epj:c44:351}, in DIS by the H1
collaboration~\cite{epj:c38:447} and in $e^+e^-$
annihilations~\cite{hep-ex-9912064p}. All the measurements are consistent.
A Monte Carlo study \cite{epj:c44:351} suggested that the measured charm
fragmentation ratios and fractions are close to those in the full $p_T(D)$
and $\eta(D)$ phase space.

The hadronisation fraction into untagged $\dz$,
needed in the next section for comparisons with theory, was: 

\begin{eqnarray*}
f^{\rm untag}(c\rightarrow \dz)
&=& f(c\rightarrow D^0)-f(c\rightarrow D^{*+})\br \\ 
&=&(\sigma^{\rm untag}(D^0) + \sigma^{\rm add}(D^{*0}))/\sigma_{gs}\\
&=&0.450 \pm 0.020 ({\rm stat.})\phantom{~}^{+0.009}_{-0.039} ({\rm syst.}) 
\phantom{~}^{+0.012}_{-0.017} ({\rm br.}){\mbox{,}}
\end{eqnarray*} 

where the equivalent phase-space treatment and $\sigma^{\rm add}(D^{*0})$
were considered as in Section~\ref{sec-gs}.

\section{Cross sections and pQCD comparisons}
\label{sec:diffxsec}

For the cross sections presented in Section~\ref{sec-xsec}
the predictions from the HVQDIS program are
$\sigma(D^{0})=7.90$ nb, $\sigma^{\rm untag}(D^{0})=5.88$ nb, $\sigma(D^{+})=2.82$ nb
and $\sigma_2(D_s^{+})=2.40$ nb, with
uncertainties around  15\%, dominated by the input PDF and the
mass of the charm quark. They are in good agreement with the data.  

\begin{figure}
\hspace*{7.0cm}{\LARGE\bf ZEUS}\\
\vspace*{-0.5cm}
\hspace*{-1.0cm}\epsfig{file=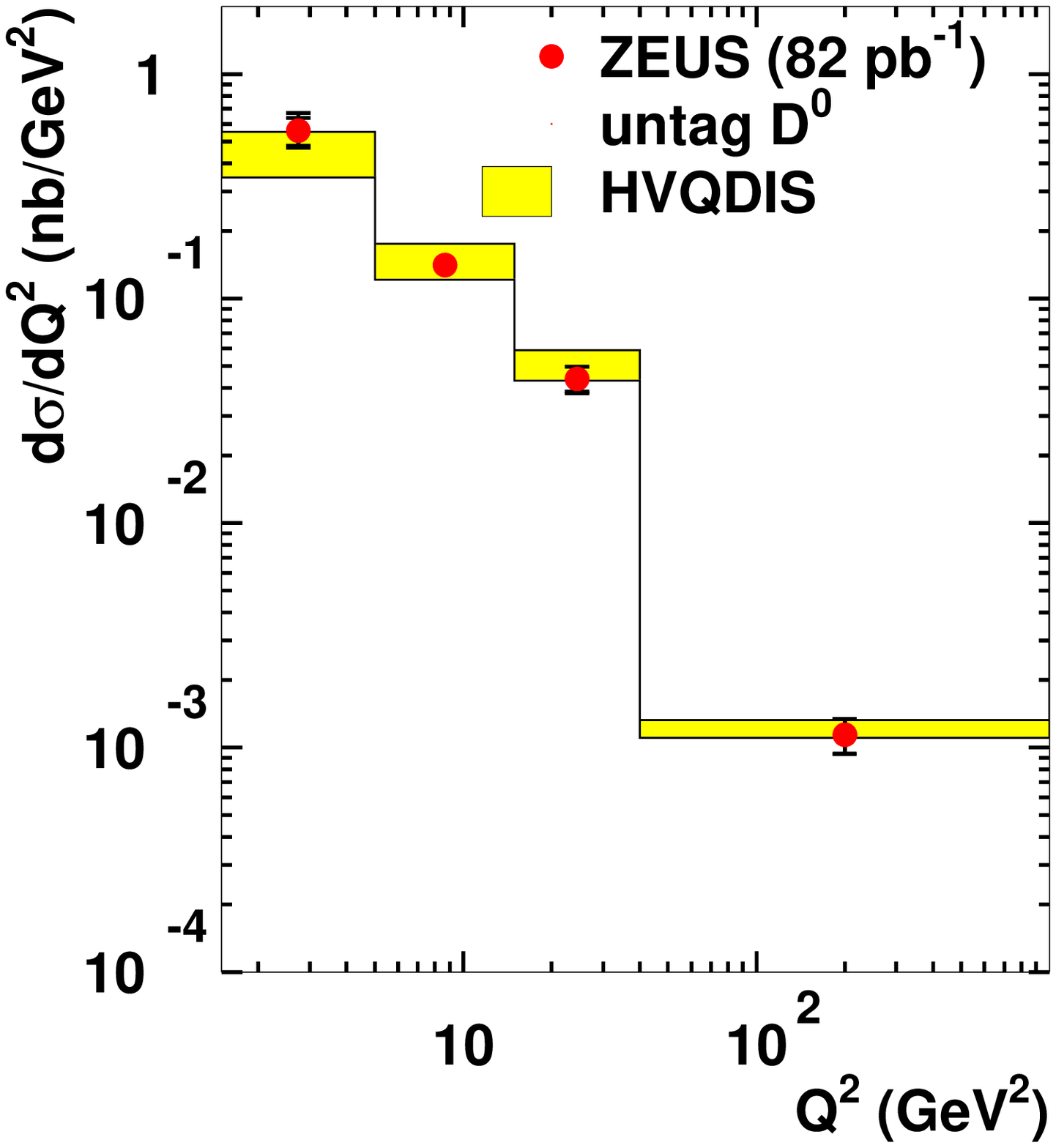,width=10cm}\hspace*{-2.0cm}\epsfig{file=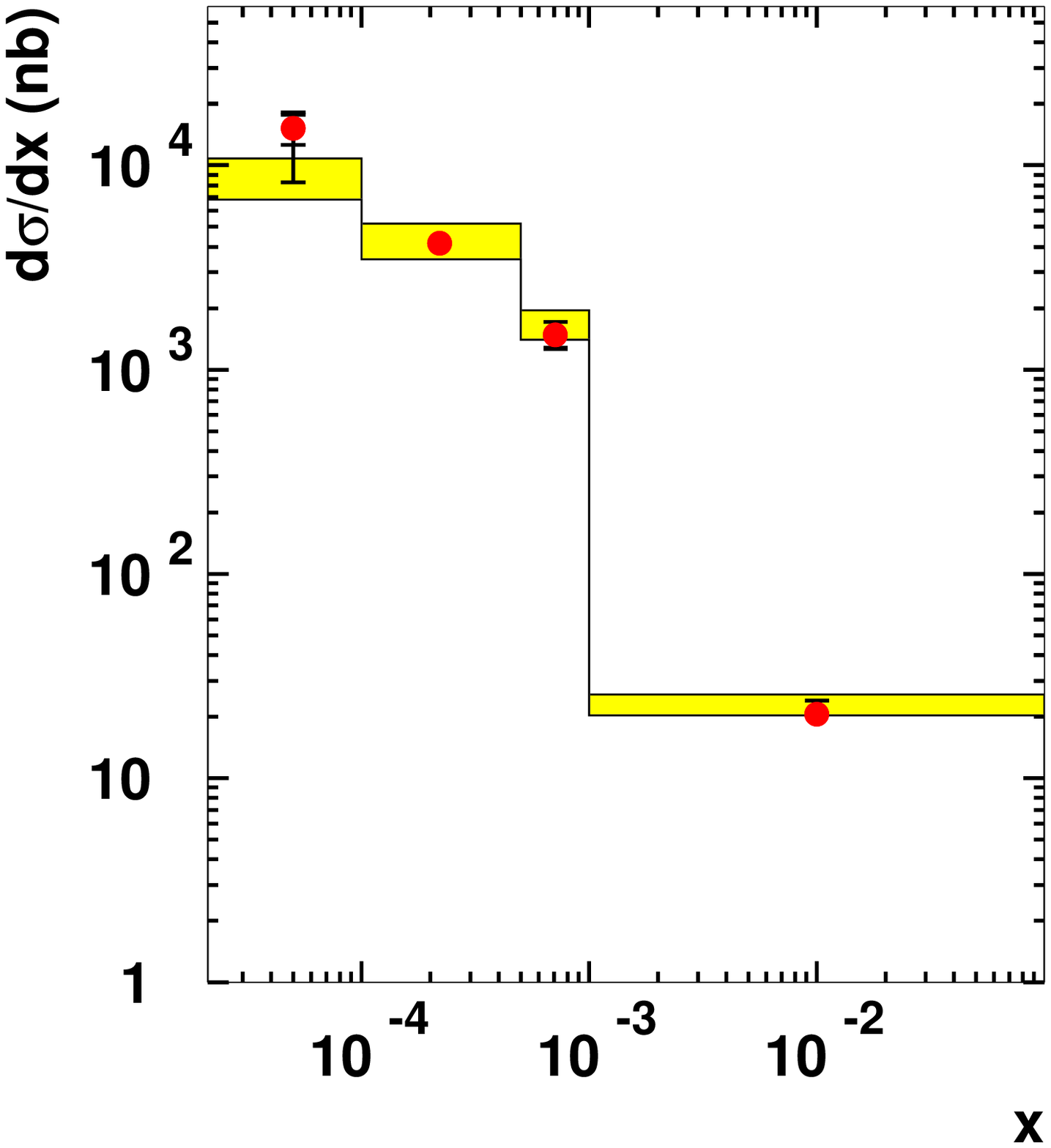,width=10cm}\\
\vspace*{-2.5cm}
\hspace*{-1.0cm}\epsfig{file=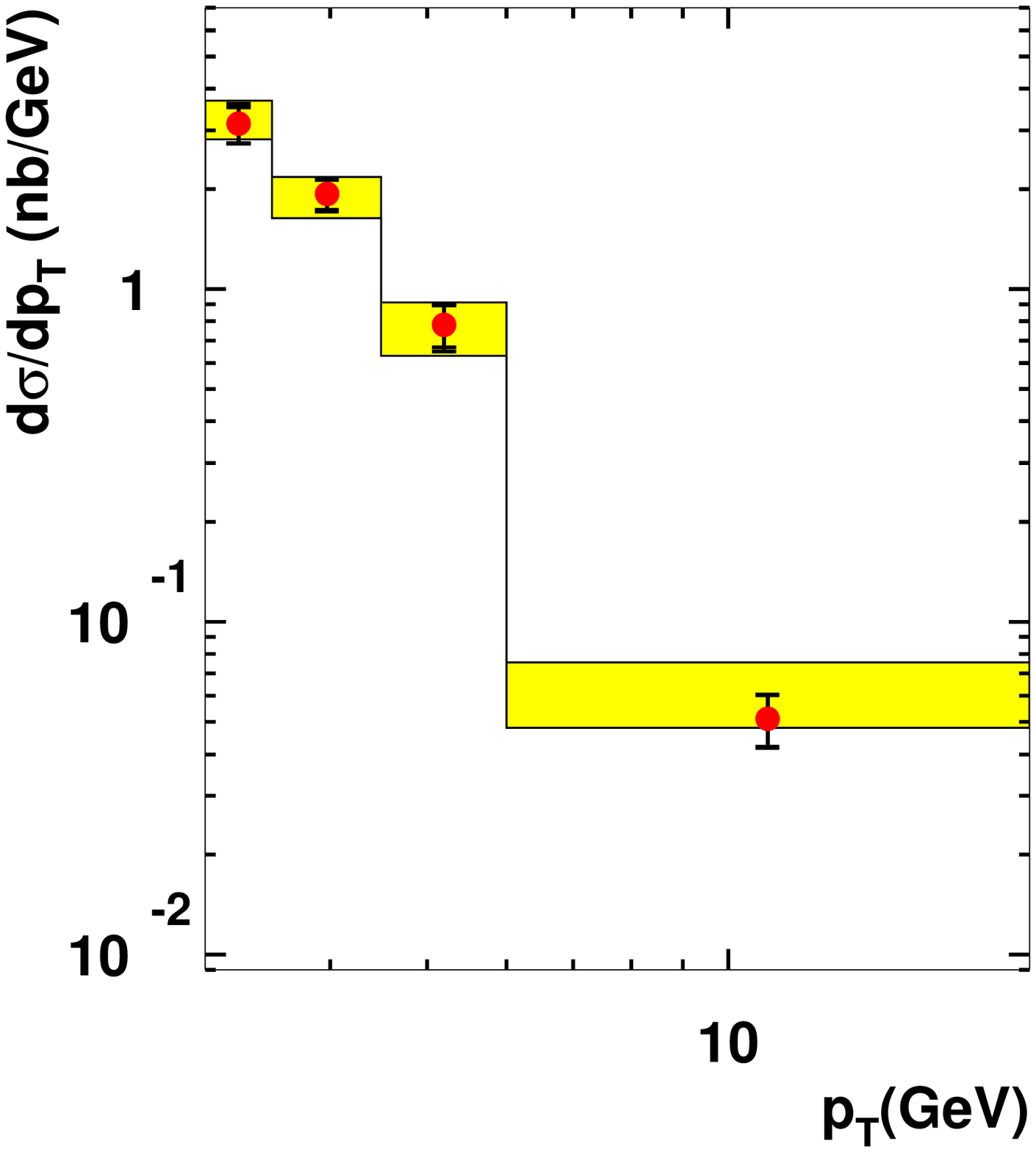,width=10cm}\hspace*{-2.0cm}\epsfig{file=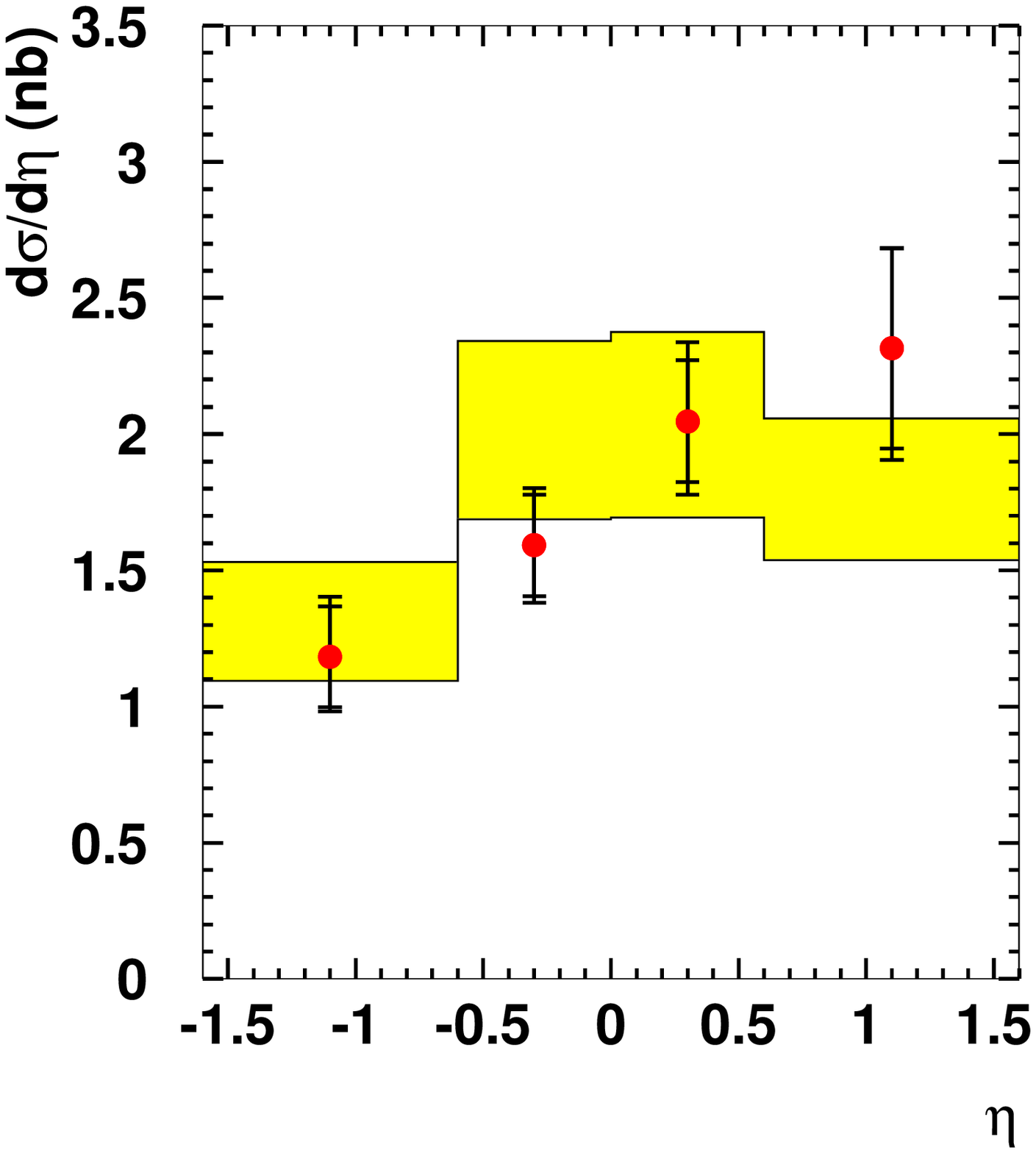,width=10cm}
\vspace*{1.0cm}
\caption
{Differential cross sections for $D^0$ not coming from $D^{*+}$ as a
function of  $Q^2$,  $x$, $p_T(D^0)$ and  $\eta(D^0)$ compared to
the NLO QCD calculation of HVQDIS. The inner error bars show the statistical
uncertainties and the outer bars show the statistical and systematic
uncertainties added in quadrature.
The lower and upper NLO QCD predictions show the estimated theoretical uncertainty
of the HVQDIS calculations.
The data points have a further 1.8\% uncertainty from the
$D^0\rightarrow K^-\pi^+$ branching ratio.
}
\label{d0xsect}
\end{figure}

\begin{figure}
\hspace*{7.0cm}{\LARGE\bf ZEUS}\\
\vspace*{-0.5cm}
\hspace*{-1.0cm}\epsfig{file=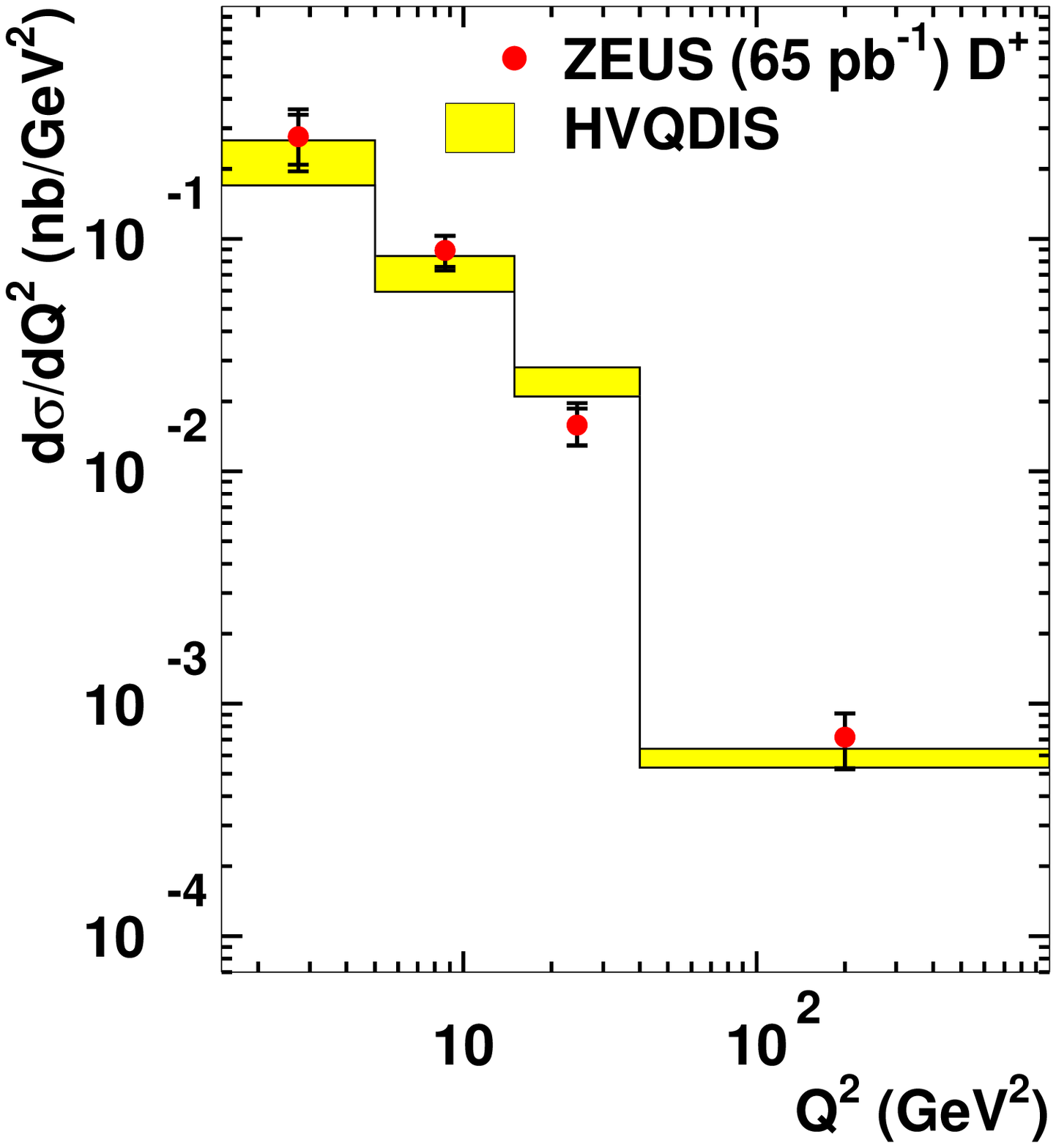,width=10cm}\hspace*{-2.0cm}\epsfig{file=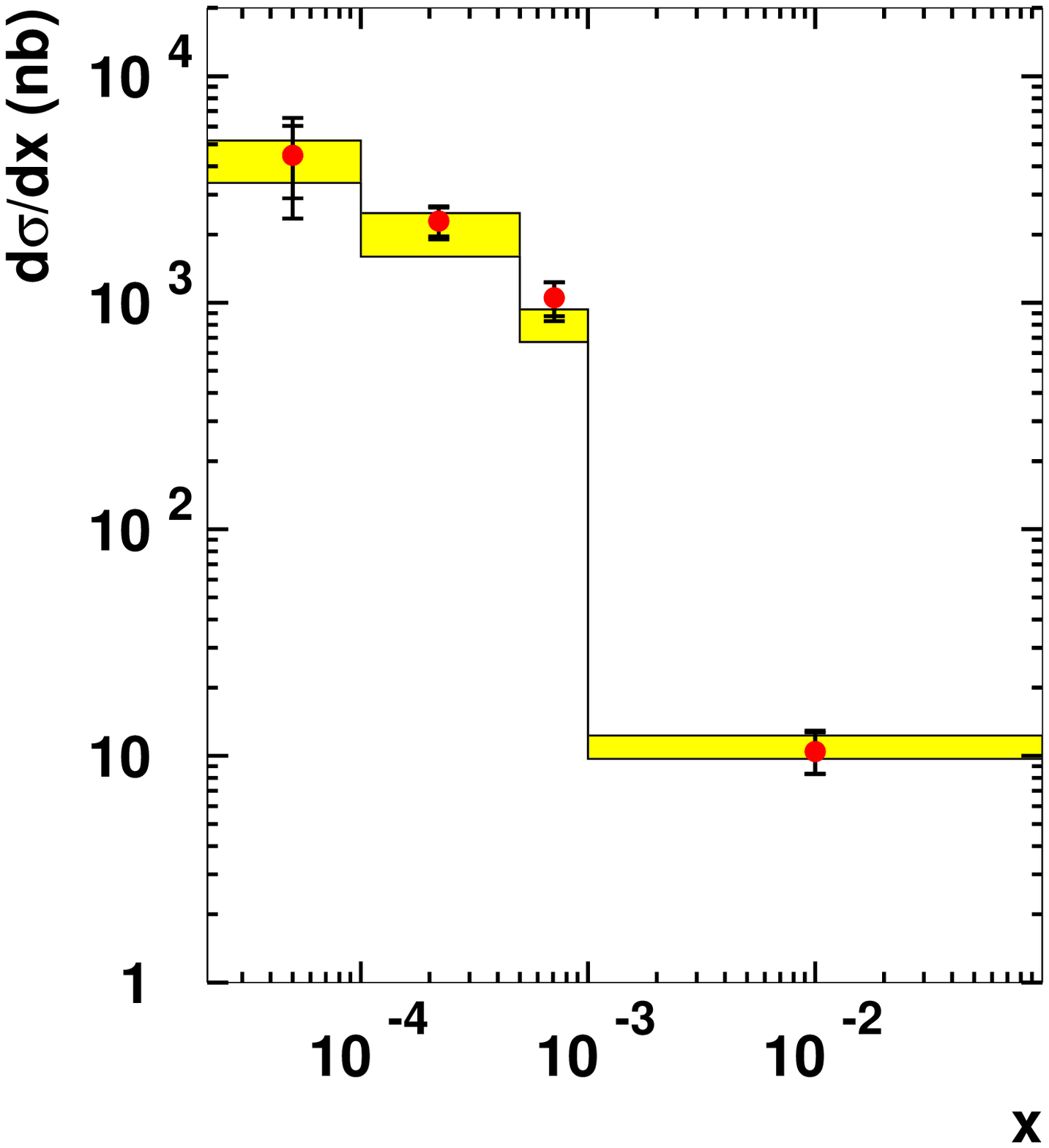,width=10cm}\\
\vspace*{-2.5cm}
\hspace*{-1.0cm}\epsfig{file=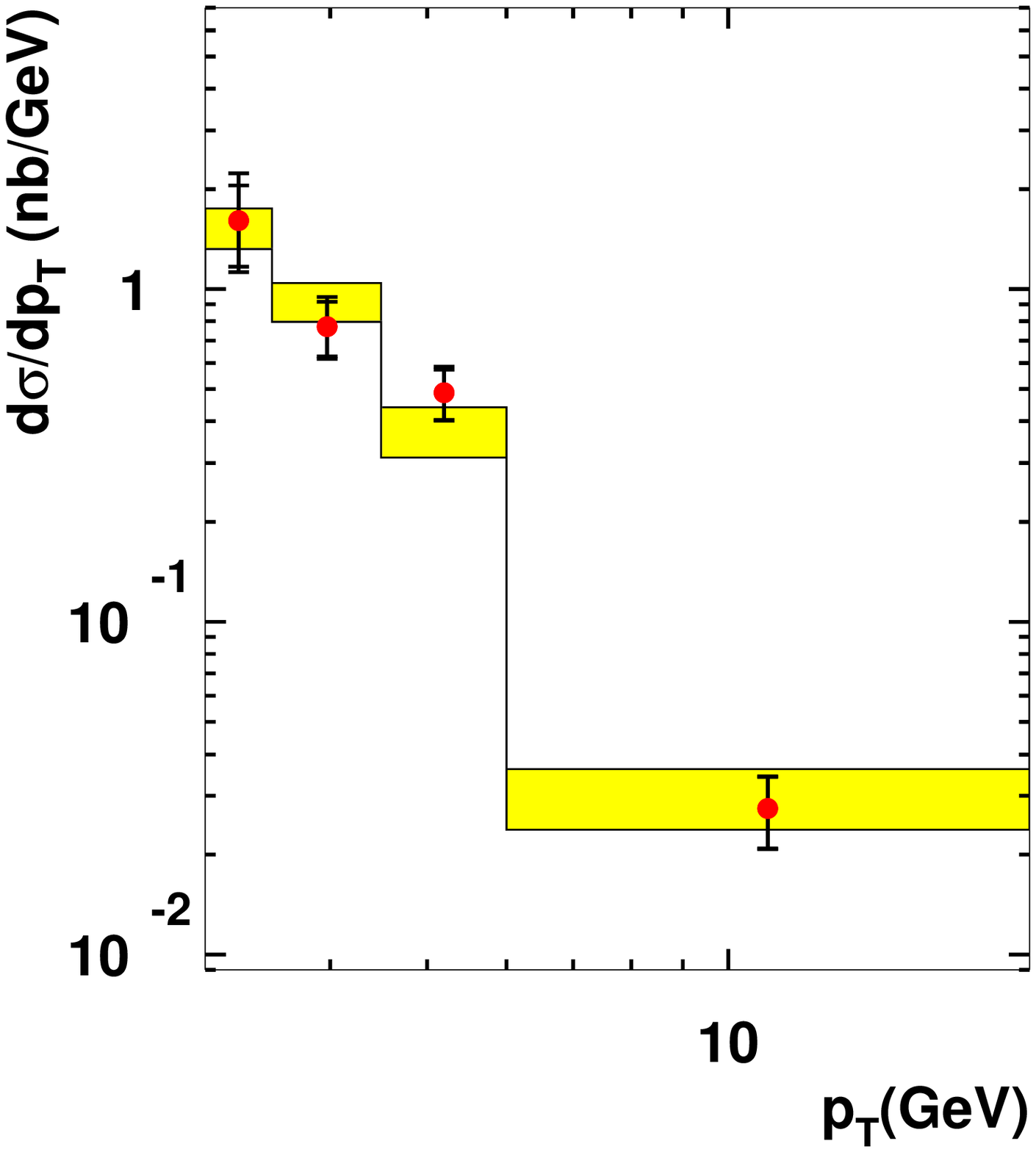,width=10cm}\hspace*{-2.0cm}\epsfig{file=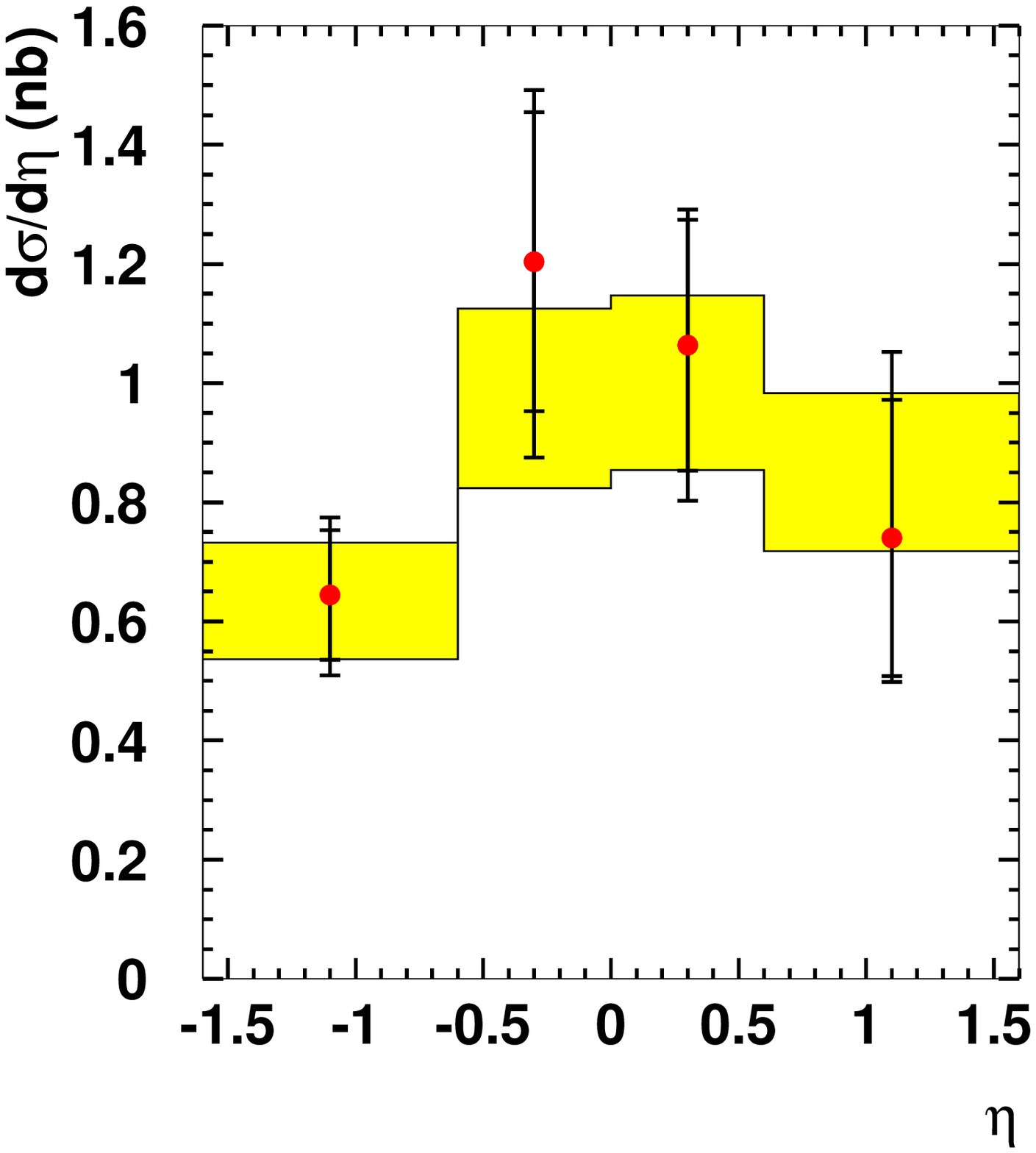,width=10cm} 
\vspace*{1.0cm}
\caption 
{Differential $D^+$ cross sections as a
function of  $Q^2$, $x$,  $p_T(D^+)$ and $\eta(D^+)$ compared to
the NLO QCD calculation of HVQDIS. The inner error bars show the statistical
uncertainties and the outer bars show the statistical and systematic
uncertainties added in quadrature.
The lower and upper NLO QCD predictions show the estimated theoretical uncertainty
of the HVQDIS calculations.
The data points have a further 3.6\% uncertainty from the
$D^+\rightarrow K^-\pi^+\pi^+$ branching ratio.
}
\label{dpmxsect}
\end{figure}

\begin{figure}
\hspace*{7.0cm}{\LARGE\bf ZEUS}\\
\vspace*{-0.5cm}
\hspace*{-1.0cm}\epsfig{file=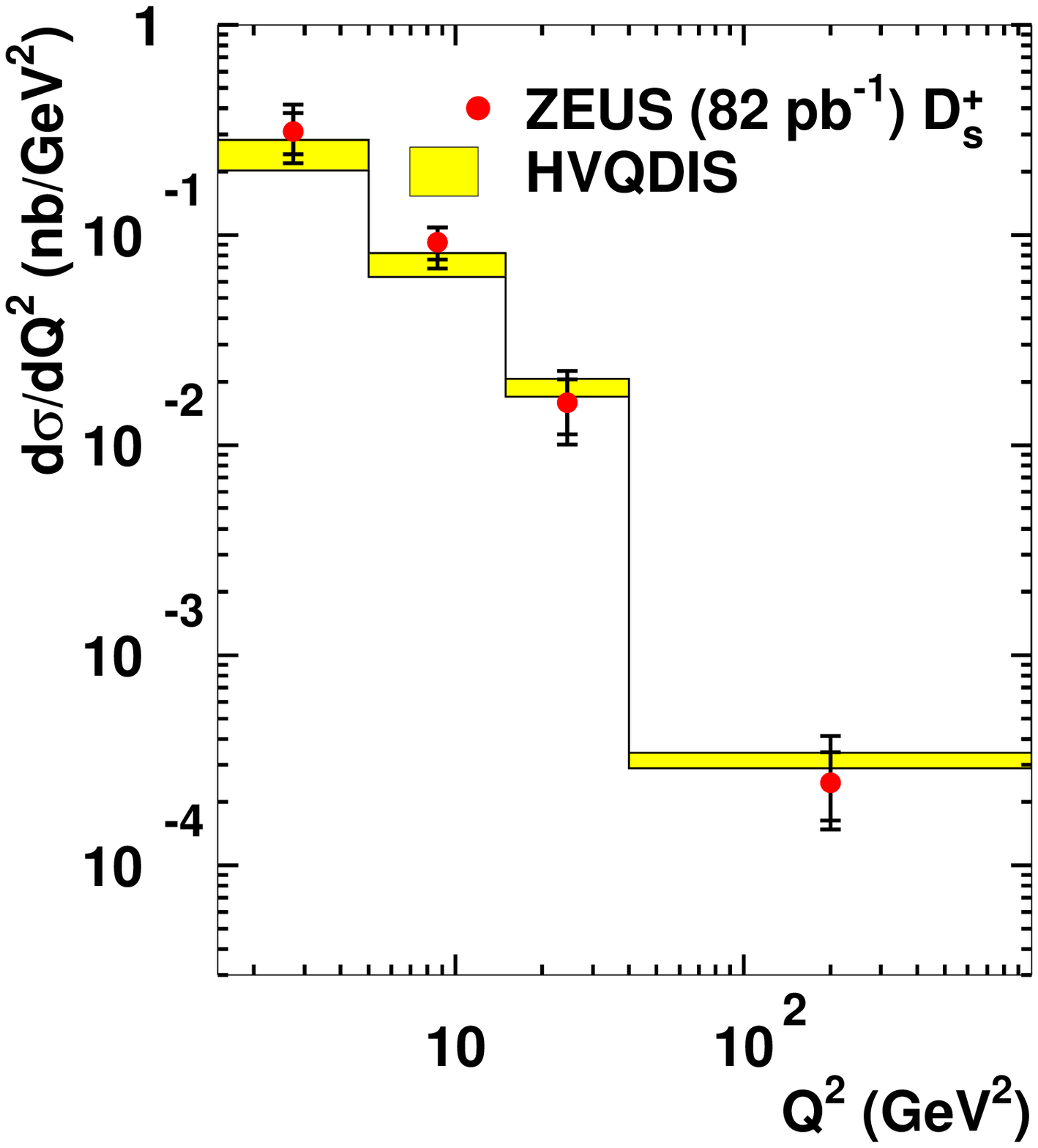,width=10cm}\hspace*{-2.0cm}\epsfig{file=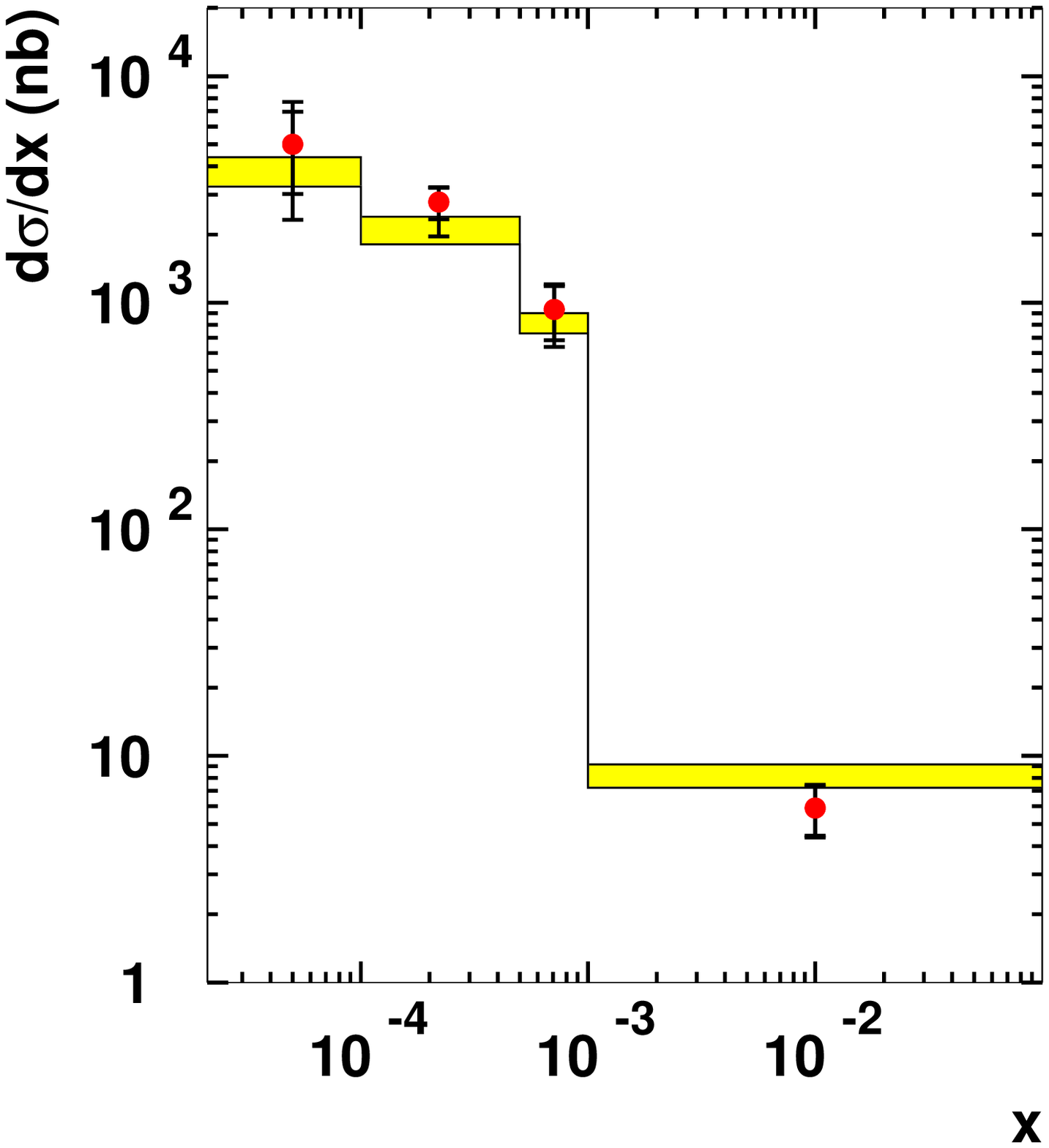,width=10cm}\\
\vspace*{-2.5cm} 
\hspace*{-1.0cm}\epsfig{file=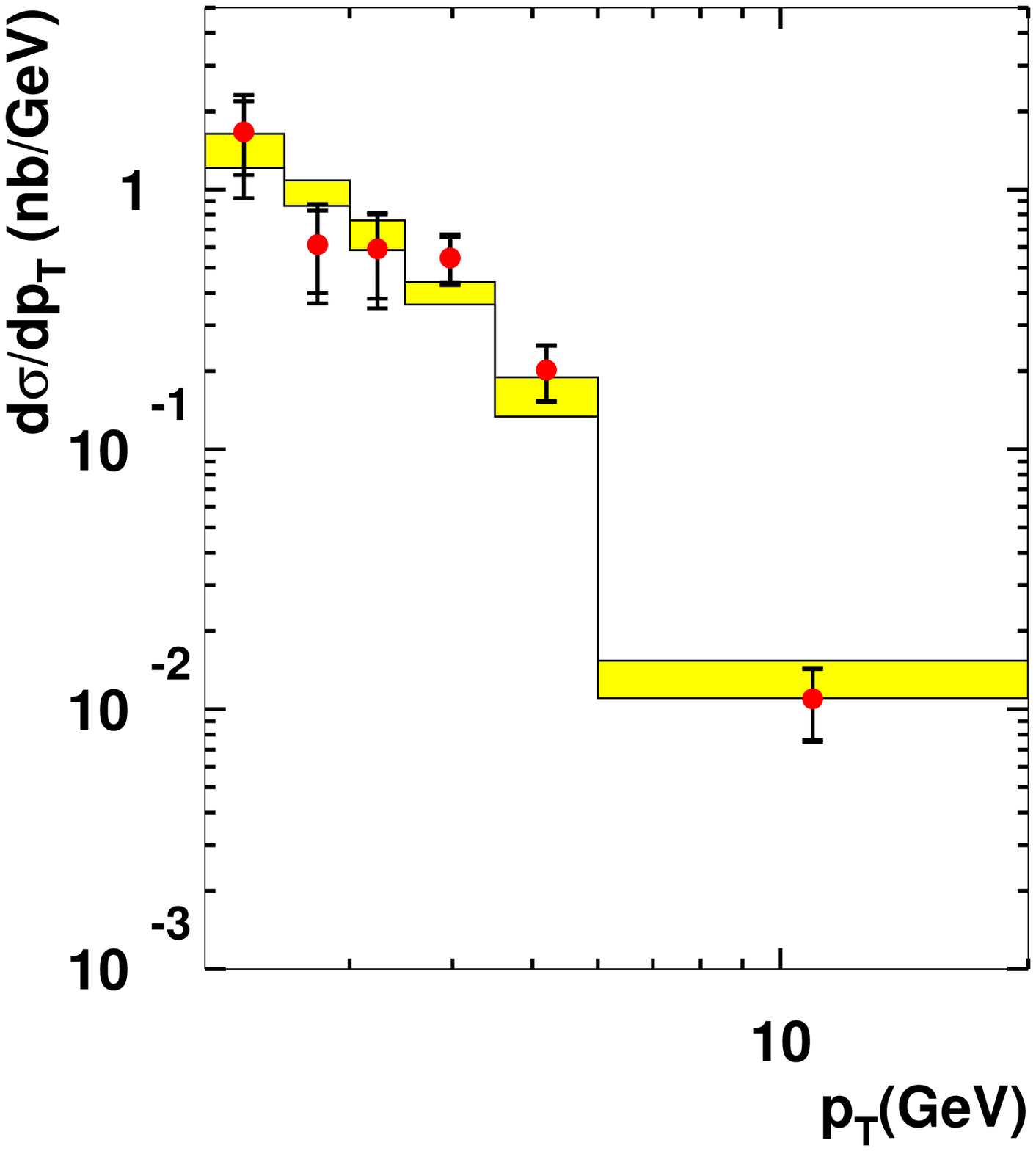,width=10cm}\hspace*{-2.0cm}\epsfig{file=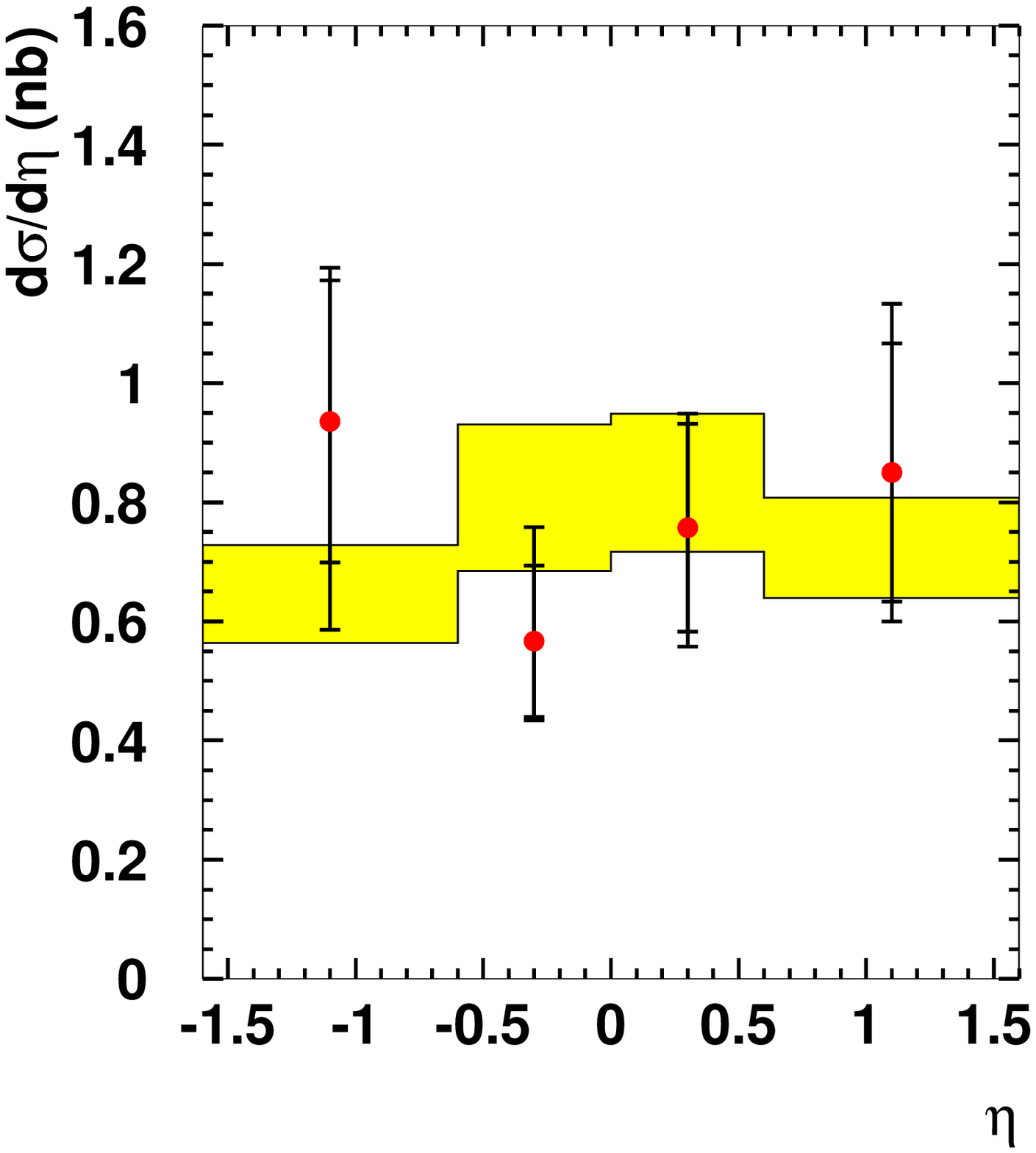,width=10cm}
\vspace*{1.0cm}
\caption
{Differential $D_s^+$ cross sections as a
function of  $Q^2$,  $x$, $p_T(D_s^+)$ and  $\eta(D_s^+)$ compared to
the NLO QCD calculation of HVQDIS. The inner error bars show the statistical
uncertainties and the outer bars show the statistical and systematic
uncertainties added in quadrature.
The lower and upper NLO QCD predictions show the estimated theoretical uncertainty
of the HVQDIS calculations.
The data points have a further 13\% uncertainty from the
$D_s^+\rightarrow  \phi\pi^+ \rightarrow K^+K^-\pi^+$ branching ratio.
}
\label{dsxsect}
\end{figure}

The differential cross sections for untagged $\dz$
(the $\dz$ mesons coming from $D^{*+}$ are already included
in the previous ZEUS publication \cite{pr:d69:012004}), 
$\dc$ and $D_s^+$  as a function of
$Q^2$, $x$, $p_T(D)$ and $\eta(D)$ are shown in
Figs.~\ref{d0xsect}, \ref{dpmxsect} and \ref{dsxsect} and given in
Tables~\ref{tab:d0dpm_bins} and \ref{tab:ds_bins}.
The cross sections in $Q^2$ and $x$ both fall by about three orders of
magnitude in the measured region. The cross-section $d\sigma/dp_T(D)$
falls by two orders of
magnitude with increasing $p_T(D)$. The cross-section $d\sigma/d\eta(D^0)$
shows a mild increase with increasing $\eta(D^0)$;
for the $\dc$ and $D_s^+$ no statistically significant dependence
with $\eta(D)$ is observed.

Figures~\ref{d0xsect}, \ref{dpmxsect} and \ref{dsxsect} show also the
corresponding NLO calculations implemented in the HVQDIS program
as well as their uncertainties (Section~\ref{sec:theo_unc}).
All the differential cross sections measured are well described
by the NLO calculation.

\section{Systematic uncertainties}
\label{sec-syst}

\subsection{Systematic uncertainties of measurements}
The systematic uncertainties of the measured cross sections
and fragmentation ratios and fractions were determined
by changing the analysis procedure and repeating all calculations. 

In the measurement of fragmentation ratios and fractions the
following groups of  systematic uncertainty sources
were considered (Table~\ref{tab:syst}):
\begin{itemize}
\item{$\{\delta_1\}$
the model dependence of the acceptance corrections was estimated using
the {\sc Herwig} MC sample;
}
\item{$\{\delta_2\}$
the uncertainty of the beauty subtraction was determined by
varying the
$b$-quark cross section by a factor of two
in the reference MC sample;
}
\item{$\{\delta_3\}$
the uncertainty of the tracking simulation was obtained
by varying all momenta by $\pm 0.3\%$
which corresponds to the uncertainty in the magnetic field;
and by changing the track momentum resolution and the angular
resolution by $^{+20}_{-10}\%$ of their values.
The asymmetric resolution variations were used since the MC signals
typically had somewhat narrower widths than observed in the data;
}

\item {$\{\delta_4\}$ the uncertainty in the CAL energy scale
was studied by varying in the MC the energy of the scattered $e^-$
by $\pm 1\%$ and the energy of the hadronic system by $\pm 3\%$;
}

\item{$\{\delta_5\}$
the uncertainties related to the signal extraction procedures were
studied as follows:}
 \begin{itemize}
  \item
   the cuts on the minimum $p_T$ for the $\pi$ and $K$ candidates
   were independently raised and lowered by 10\% from their nominal values,
  \item
    the cut on the minimum $p_T$ for the $\pi_s$
   was raised and lowered by $0.02$ GeV
   (for $\sigma^{\rm tag}(D^{0})$, $\sigma^{\rm untag}(D^{0})$, $\sigma^{\rm add}(D^{*+})$),
  \item
   the $\Delta M$ signal region
   was widened symmetrically by $0.003$ GeV
   (for $\sigma^{\rm tag}(D^{0})$, $\sigma^{\rm untag}(D^{0})$, $\sigma^{\rm add}(D^{*+})$),
  \item
   the $M(K\pi)$ signal region was widened and narrowed symmetrically by $0.01$ GeV
   (for $\sigma^{\rm add}(D^{*+})$),
  \item
   the wrong-charge background normalisation region was changed to $0.152 < \Delta M < 0.168$
   (for $\sigma^{\rm add}(D^{*+})$);
 \end{itemize}
\item{$\{\delta_6\}$
the uncertainties of the luminosities of the $e^-p$ ($\pm1.8\%$) and $e^+p$
($\pm2.25\%$) data samples were included, taking into account their
correlations;
}
\item{$\{\delta_7\}$
the uncertainty in the estimate of $\sigma(\Lambda_c^+)$
(see Section~\ref{sec-ff});
}
\item{$\{\delta_8\}$
the uncertainty in the rate of the charm-strange baryons
(see Section~\ref{sec-ff});
}
\end{itemize}

Contributions from
the different systematic uncertainties were calculated and added 
in quadrature separately for positive and negative variations.
Correlated systematic uncertainties largely cancel
in the calculation of the fragmentation ratios and fractions.

For the total and  differential cross-section measurements
discussed in Section \ref{sec:diffxsec} and those used
for the extraction of $F_2^{c\bar{c}}$ (Section \ref{sec:extract}),
further sources of systematics were studied
\cite{pr:d69:012004,epj:c12:35,thesis:zabrana:2006}, $\{\delta_9\}$:

\begin{itemize}

\item the cut on $y_e$ was changed to $y_{e}\> \leq\>  0.90$;

\item the cut on $y_{\mathrm{JB}}$ was changed to $y_{\mathrm{JB}}\> \geq\> 0.03$;

\item the cut on $\delta$ was changed to $42 \>  \leq \> \delta\>  \leq\>  70$ GeV;

\item the cut on  $|Z_{\rm vertex}|$ was changed to $|Z_{\rm vertex}| < 45$ cm;

\item the cut on $E_{e^\prime}$ was changed to $E_{e^\prime} > 11$~GeV;

\item the excluded region for the impact position of the scattered
      electron in the RCAL was increased by 1~cm in each direction;

\item the electron method was used, except
      for cases when the scattered electron
      track was reconstructed by the CTD. In the latter case, the $\mathrm{DA}$
      method, which has the best resolution at high $Q^2$, was used.
  
\end{itemize}
   
These estimations were  made in each bin in which the differential cross sections were
measured. In addition, for the lowest $x$ bin of the differential cross section of
untagged $D^0$, the systematic error accounted also for instabilities in the signal
extraction, not encountered in any other bin. The overall systematic uncertainty was
determined by adding the individual uncertainties in quadrature. Typically $\delta_9$
was below $ 4\%$. The uncertainty on the luminosity measurement was not included in the
systematic uncertainties for the differential cross sections.
   
\subsection{Uncertainties on theoretical predictions}
\label{sec:theo_unc}
The NLO QCD predictions for $D$ meson production are affected by
the systematic uncertainties listed below.
Typical values are quoted for the total cross section:

\begin{itemize}

\item the ZEUS PDF uncertainties were propagated from the experimental
      uncertainties of the fitted data ($\pm 5\%$). As an alternative
      parametrisation in the FFNS, the CTEQ5F3 PDF was used in HVQDIS
      with a charm mass of 1.3 GeV ($-2\%$);

\item the charm mass was changed simultaneously in the PDF fit and in HVQDIS by
      $\mp 0.15$~GeV $\left( ^{+8}_{-8}\% \right)$.
       The largest effect was at low $p_T(D)$;

\item the scale was changed to $2\sqrt{Q^2+4m_c^2}$ and to
      ${\rm max}(\sqrt{Q^2/4+m_c^2},2m_c)$
      $\left(^{+5}_{-6}\% \right)$;

\item  the {\sc Jetset} fragmentation as implemented in the
       previous analyses~\cite{pr:d69:012004,epj:c12:35}
       was used instead of the Peterson fragmentation ($+5\%$ to $+20\%$).
       The largest deviations were observed for $D^0$ and $D^+$ particles
       at the lowest $Q^2$ and $x$.

\end{itemize}

\section{Extraction of \ftwoccb}
\label{sec:extract}

The open charm contribution, \ftwoccb, to the proton structure-function \ftwo\
can be defined in terms of the inclusive double-differential $c\bar{c}$ cross
section in $x$ and \qsq\ by

$$
\frac{d^2\sigma^{c\bar{c}} (x, Q^2)}{dxdQ^2} =
\frac{2\pi\alpha^2}{x Q^4}
\{ [1+(1-y)^2] F_2^{c\bar{c}}(x, Q^2) - y^2 F_L^{c\bar{c}}(x, Q^2) \} .
$$

In this paper, the $c\bar{c}$ cross section is obtained by measuring the
untagged $D^0$, $D^+$ and $D_s^+$
production cross sections and employing the measured hadronisation fractions
$f(c \rightarrow D)$.  Since only
a limited kinematic region in $p_T(D)$ and $\eta(D)$ is accessible,
a prescription for extrapolating to the full kinematic phase space is needed.

As reported in Section~\ref{sec:diffxsec}, the measured differential cross-sections are well
described in the probed kinematic region. Therefore the following relation
was used to extract \ftwoccb:

\begin{equation}
F_{2}^{c\bar{c}}(x_i, Q^2_i) = \frac{\sigma_{i,\rm meas}(ep \rightarrow D X)}
                                      {\sigma_{i,\rm theo}(ep \rightarrow D X)}
                                      F_{2,\rm theo}^{c\bar{c}}(x_i, Q^2_i),
\label{eq:f2cc}
\end{equation}

where $\sigma_{i,\rm meas}$ is the cross section in the bin $i$ in the
measured region of $p_T(D)$ and $\eta(D)$ and $\sigma_{i,\rm theo}$ is
the corresponding cross section evaluated with HVQDIS.
 The value of $\ftwoccb_{\rm theo}$ was calculated in FFNS from the NLO
coefficient functions~\cite{pr:d67:012007,*misc:www:zeus2002}
using the same values of parameters as in the calculation of $\sigma_{i,\rm theo}$.
The cross sections $\sigma_{i,\rm meas}(ep \rightarrow D X)$ were measured
in bins of  $Q^2$ and $y$ (Table~\ref{tab:dd_bins})
and $F_2^{c\bar{c}}$ was quoted at representative
$Q^2$ and $x$ values for each bin (Table~\ref{tab:f2c_bins}).
The $F_2^{c\bar{c}}$ measurements obtained from each $D$ meson were combined 
into a single set of measurements; the result is also shown in
Table~\ref{tab:f2c_bins}. 

The extrapolation factors from the measured $p_T(D)$ and $\eta(D)$ ranges to
the full phase space, as estimated using HVQDIS, were
between 17 at low $Q^2$ and 2.5  at high $Q^2$ for the $D^0$ and $D^+$
measurements.  For the $D_s^+$, the lower
$p_T$ requirement leads to lower extrapolation factors between
$5.6 $ and  $1.9 $. They are all shown in Table~\ref{tab:f2c_bins}.
The uncertainty from the branching ratios was estimated by 
changing each branching ratio independently in the calculation
by $\pm 1$ standard deviation and 
adding in quadrature the resulting variations of
$F_2^{c\bar{c}}$ $(^{+2.7}_{-4.1}\%)$.

The following uncertainties of the extrapolation 
prescription of Eq.~(\ref{eq:f2cc}) have been evaluated:

\begin{itemize}

\item using {\sc Jetset} instead of the Peterson
      fragmentation yielded changes of $\approx +28\%$, $+15\%$ and $+5\%$ 
      for the data points at the lowest, middle and largest $Q^2$ ranges,
      respectively; 

\item changing the charm mass by $\pm 0.15$~GeV consistently in the HVQDIS
      calculation and in the calculation of $F_{2, \rm theo}^{c\bar{c}}$ led to
      differences in the extrapolation of $\pm 5\%$  at low $x$, low $Q^2$;
     the value
      decreases rapidly at higher $x$ and higher $Q^2$;

\item the uncertainty in the ZEUS NLO PDF fit led to uncertainties
      in the extracted values of $F_2^{c\bar{c}}$ typically less than
      $1\%$;

\item the extrapolation factors were evaluated using the CTEQ5F3 proton PDF
      yielding differences compared to the nominal factors of
      $\approx +10\%$, $+6\%$ and $+1\%$
       for the lowest, middle and largest $Q^2$ ranges, respectively.

\end{itemize}

\begin{figure}
\hspace*{7.5cm}{\Large\bf ZEUS}\\
\epsfig{file=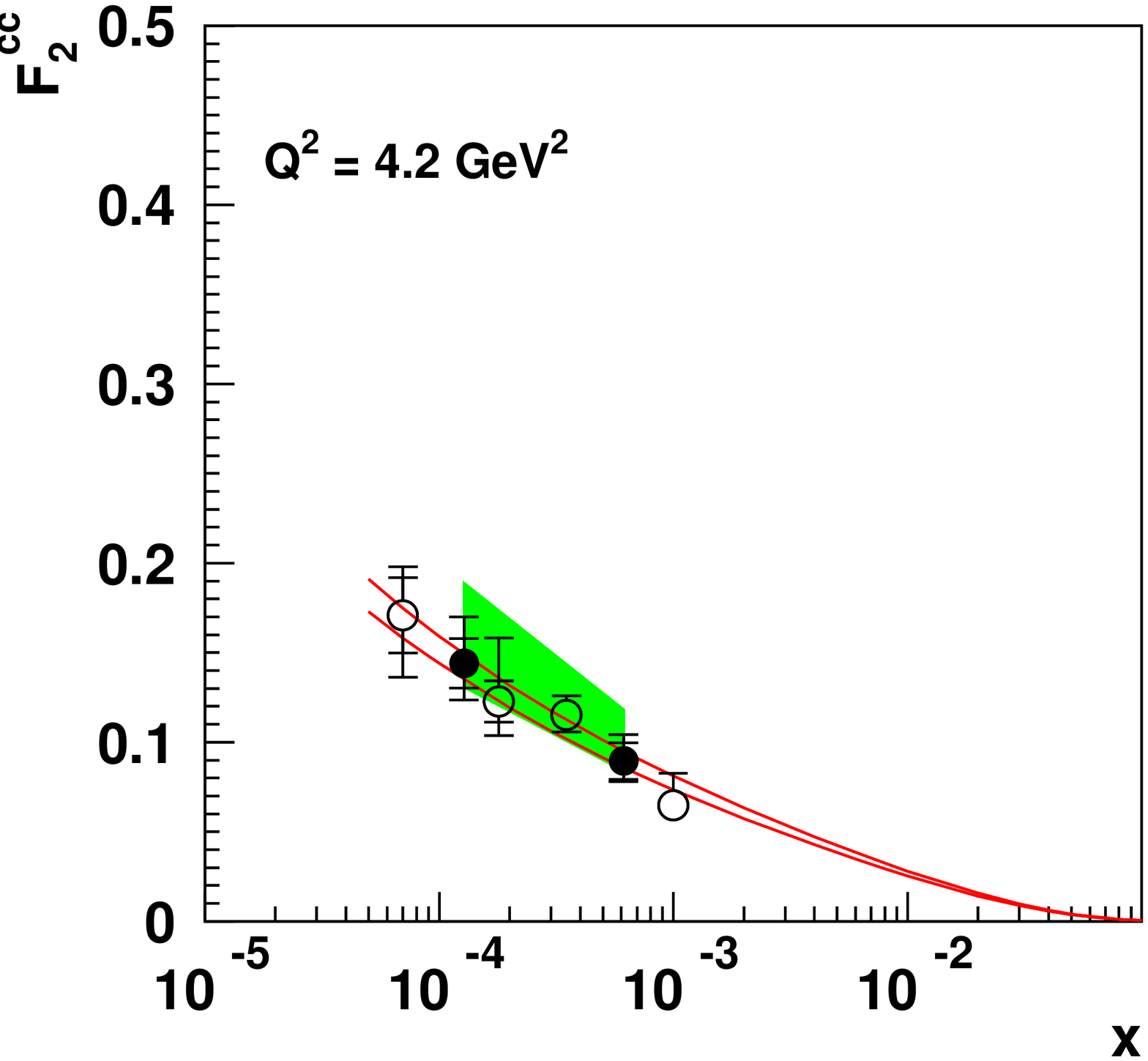,height=8cm}\epsfig{file=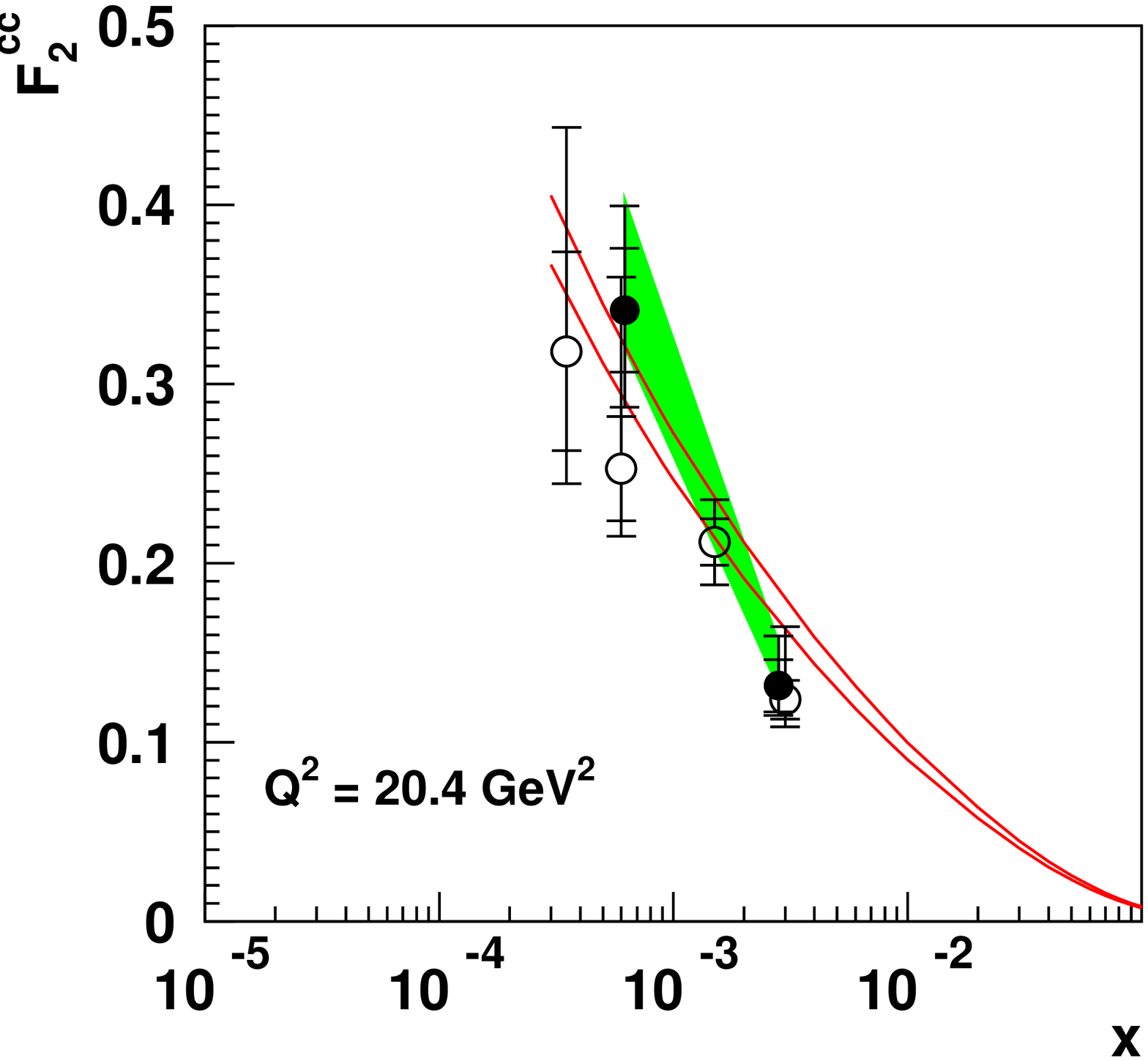,height=8cm}\\
\epsfig{file=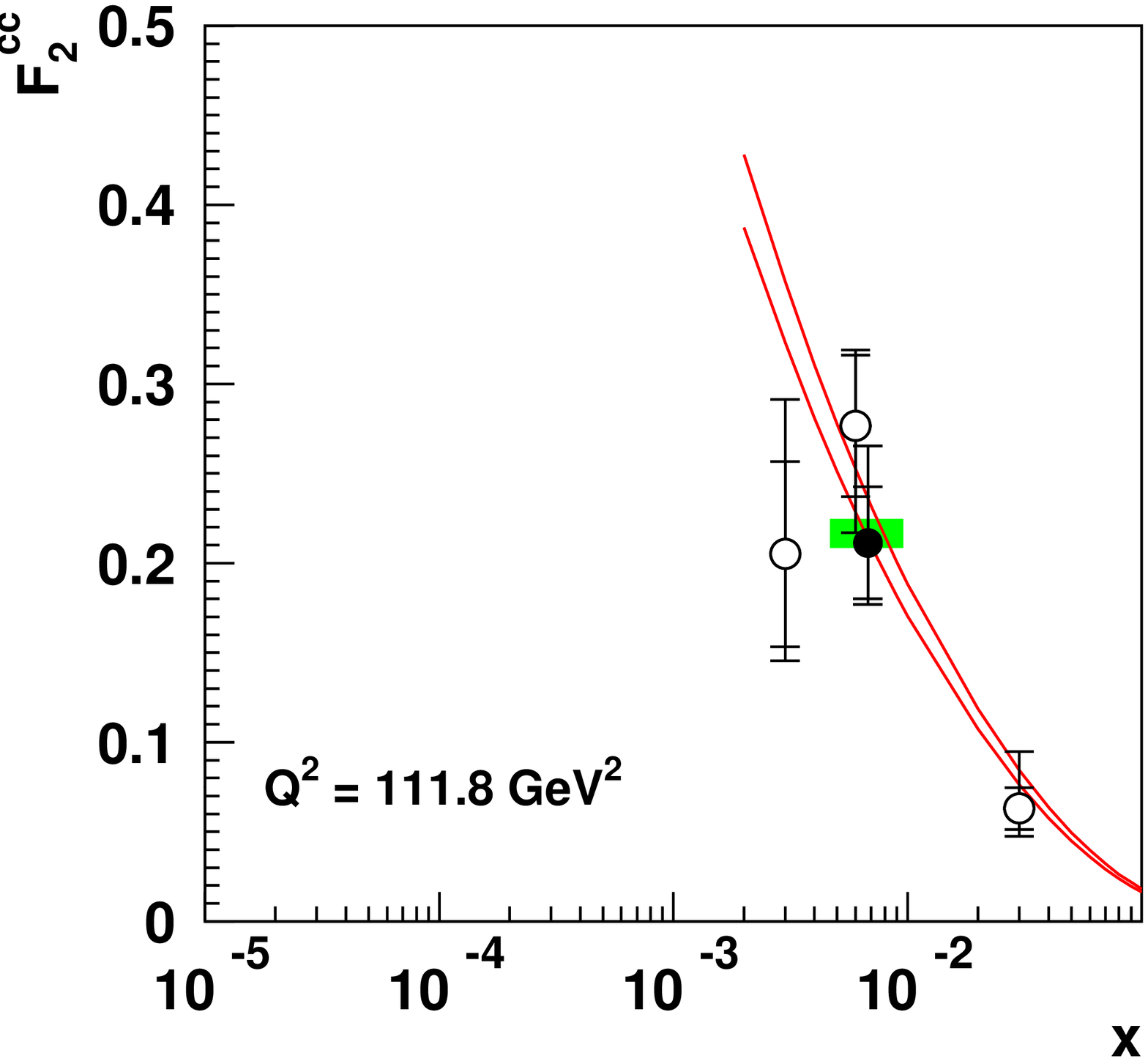,height=8cm}\epsfig{file=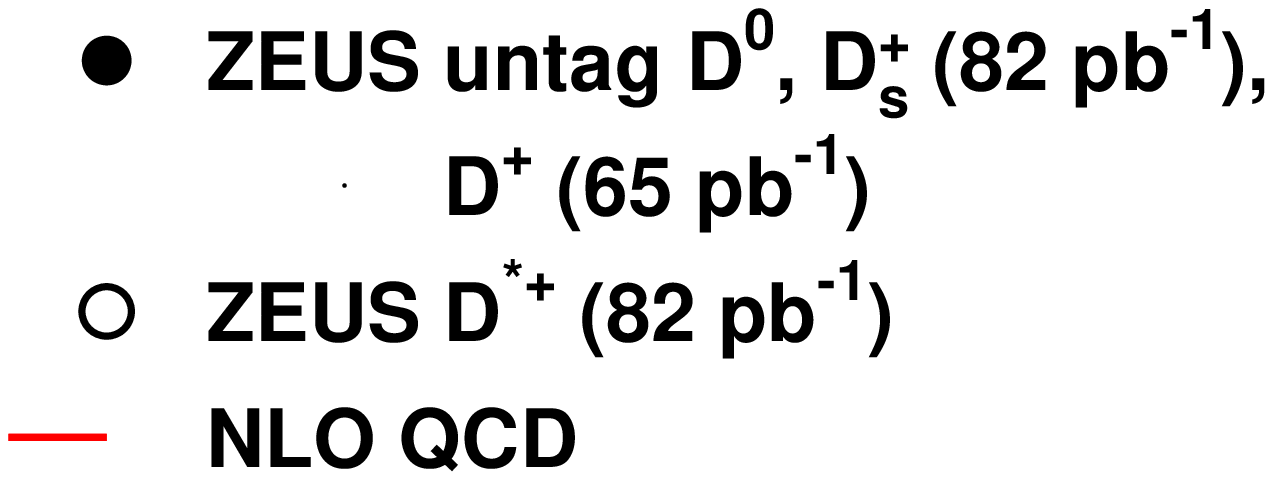,height=8cm}
\caption
{The measured $F_2^{c \bar{c}}$  as a function of $x$ for three  $Q^2$ bins.
The current data are compared with the previous
ZEUS measurement~\protect\cite{pr:d69:012004}. The data are shown with statistical uncertainties
(inner bars) and statistical and systematic uncertainties added in quadrature
(outer bars).
All measured $F_2^{c\bar{c}}$ values have a further +2.7\% -4.1\%
uncertainty coming from the current experimental uncertainty from the
$D^0\rightarrow K^-\pi^+$, $D^+\rightarrow K^-\pi^+\pi^+$ and
$D_s^+\rightarrow  \phi\pi^+ \rightarrow K^+K^-\pi^+$
branching ratios and the $f(c\rightarrow \Lambda_c^+)$ value.
 The shaded band corresponds to the
 estimated theoretical uncertainty in the extrapolation.
The lower and upper
 curves show the ZEUS NLO QCD fit~\protect\cite{pr:d67:012007,*misc:www:zeus2002}
uncertainty propagated from the experimental uncertainties of the fitted data.
}

\label{f2charm_vs_x}
\end{figure}

The combined  $F_2^{c\bar{c}}$ measurements are 
shown in Fig.~\ref{f2charm_vs_x}.
The quadratic addition of the  extrapolation uncertainties is shown  
independently as a band.
Also shown in Fig.~\ref{f2charm_vs_x} is the previous
 measurement~\cite{pr:d69:012004} and the ZEUS NLO QCD fit.
The two sets of data are consistent\footnote{The previous
data were measured at $Q^2= 4$, $18$ and $130$~GeV$^2$ and 
  have been shifted to  $Q^2= 4.2$, $20.4$ and $111.8$~GeV$^2$ using the
ZEUS NLO QCD fit.}.
The prediction describes the data well for all $Q^2$ and $x$.
The uncertainty on the theoretical prediction is that
from the PDF fit propagated from the experimental 
uncertainties of the fitted data.

\section{Summary and conclusions}
The production of the charm mesons
$\dsp$, $\dz$, $\dc$ and $\dssp$
has been measured with the ZEUS detector
in the kinematic range
 $ 1.5 <Q^2<1000\gev^2$, $0.02<y<0.7$,
$p_T(\dsp,\dz,\dc)>3\gev$, $p_T(\dssp)>2\gev$
and $|\eta(D)|<1.6$.

 The cross sections have
been used to determine the charm fragmentation ratios and fractions.
The ratio of neutral to charged $D$-meson production,
$R_{u/d}$, is compatible with unity, i.e.$\:$it is
consistent with isospin invariance, which implies that $u$ and $d$ quarks are
 produced equally in charm fragmentation.
The  strangeness-suppression factor in charm fragmentation, $\gamma_s$,
was measured to be about 20\%.
The fraction of charged $D$ mesons produced in a vector state, $P^d_{\rm v}$,
 was found to be smaller than the naive spin-counting prediction of $0.75$.
The fraction of $c$ quarks hadronising as $\dsp$, $\dz$, $\dc$ and  $\dssp$
mesons have been calculated.
The measured $R_{u/d}$, $\gamma_s$, $P^d_{\rm v}$ and open charm
fragmentation fractions are consistent with those
obtained in charm photoproduction and in $e^+e^-$ annihilation.
These measurements generally support the hypothesis that fragmentation
proceeds independently of the hard sub-process.

The measured $\dz$, $\dc$ and $\dssp$ differential cross sections
were compared to the predictions of NLO QCD with  the proton PDFs
extracted from inclusive DIS data.  A good description was found.

The double-differential cross section in $y$ and $Q^2$ has been
used to extract the open charm contribution to $F_2$, by
using the NLO QCD calculation to extrapolate outside the
measured $p_T(D)$ and $\eta(D)$ regions. The $F_2^{c\bar{c}}$ values obtained
from the different $D$ mesons agree with previous results where
a $D^{*+}$ meson was tagged.

\section*{Acknowledgments}
The strong support and encouragement of the DESY Directorate have been
invaluable, and we are much indebted to the HERA machine group for their
inventiveness and diligent efforts. The design, construction and installation
of the ZEUS detector have been made possible by the ingenuity and dedicated
efforts of many people from inside DESY and from the home institutes who are
not listed as authors. Their contributions are acknowledged with great
appreciation.


\newpage

\providecommand{\etal}{et al.\xspace}
\providecommand{\coll}{Coll.\xspace}
\catcode`\@=11
\def\@bibitem#1{%
\ifmc@bstsupport
  \mc@iftail{#1}%
    {;\newline\ignorespaces}%
    {\ifmc@first\else.\fi\orig@bibitem{#1}}
  \mc@firstfalse
\else
  \mc@iftail{#1}%
    {\ignorespaces}%
    {\orig@bibitem{#1}}%
\fi}%
\catcode`\@=12

\newpage

\vspace*{2.cm}
\begin{table}[hbt]
\begin{center}
\begin{tabular}{|c|c|}\hline
Decay mode & Branching ratio \\
\hline
\hline
$D^{0}\rightarrow K^{-}\pi^{+}$      & $0.0380 \pm 0.0007$ \\ \hline
$\dsp \rightarrow \dz \pi^+_s $      & $0.6770 \pm 0.0050$ \\ \hline
$D^+\rightarrow K^{-}\pi^{+}\pi^{+}$ & $0.0951 \pm 0.0034$ \\ \hline
$D_s^{+}\rightarrow \phi \pi^{+} \rightarrow K^{+}K^{-} \pi^{+}$    & $0.0216 \pm 0.0028$ \\ \hline
\end{tabular}
\caption{Branching ratios of the $D$-meson decay modes~\protect\cite{jp:g33:1}.}
\label{tab:branching}
\end{center}
\end{table}
%
%
\newpage
\clearpage
\begin{table}[hbt]
\begin{center}
\begin{tabular}{|c|c|} \hline
& $R_{u/d}$ \\
\hline
\hline
ZEUS (DIS)
&
$1.22\pm0.11({\rm stat.})^{+0.05}_{-0.02}({\rm syst.})\pm 0.03({\rm br.})$
\\
\hline
ZEUS ($\gamma p$)~\cite{epj:c44:351}
&
$1.100\pm0.078({\rm stat.})^{+0.038}_{-0.061}({\rm syst.})^{+0.047}_{-0.049}({\rm br.})$
\\
\hline
combined $e^+e^-$ data
& $1.020\pm0.069({\rm stat.\oplus syst.})^{+0.045}_{-0.047}({\rm br.})$ \\
\hline
H1 (DIS)~\cite{epj:c38:447}
& $1.26\pm0.20({\rm stat.})\pm0.11({\rm syst.})\pm0.04({\rm br.\oplus theory})$ \\
\hline
\end{tabular}
\caption{
The ratio of neutral to charged $D$-meson production rates, $R_{u/d}$.
The $e^+e^{-}$ values are taken from ~\protect\cite{epj:c44:351}; they are an
update of the compilation in~\protect\cite{hep-ex-9912064p} using the
branching-ratio values of~\protect\cite{pl:b592:1}. The measurements in this
paper, ZEUS (DIS), took the values for all the branching ratios involved
from~\protect\cite{jp:g33:1}, the rest of the quoted measurements took
them from~\protect\cite{pl:b592:1}.
}
\label{tab:rud}
\end{center}
\end{table}
\vspace*{2.cm}
\begin{table}[hbt]
\begin{center}
\begin{tabular}{|c|c|} \hline
& $\gamma_s$ \\
\hline
\hline
ZEUS (DIS)
& $0.225 \pm 0.030({\rm stat.})^{+0.018}_{-0.007}({\rm syst.})^{+0.034}_{-0.026}({\rm br.})$ \\
\hline
ZEUS ($\gamma p$)~\cite{epj:c44:351}
& $0.257 \pm 0.024({\rm stat.})^{+0.013}_{-0.016}({\rm syst.})^{+0.078}_{-0.049}({\rm br.})$ \\
\hline
ZEUS ($\gamma p$)~\cite{pl:b481:213}
& $0.27 \pm 0.04({\rm stat.})^{+0.02}_{-0.03}({\rm syst.})\pm 0.07({\rm br.})$ \\
\hline
combined $e^+e^-$ data
& $0.259\pm0.023({\rm stat.\oplus syst.})^{+0.087}_{-0.052}({\rm br.})$ \\
\hline
H1 (DIS)~\cite{epj:c38:447}
& $0.36\pm0.10({\rm stat.})\pm0.01({\rm syst.})\pm0.08({\rm br.\oplus theory})$ \\
\hline
\end{tabular}
\caption{
The strangeness-suppression factor in charm fragmentation, $\gamma_s$.
The $e^+e^{-}$ values are taken from ~\protect\cite{epj:c44:351}; they are an
update of the compilation in~\protect\cite{hep-ex-9912064p} using the
branching-ratio values of~\protect\cite{pl:b592:1}. The measurements in this
paper, ZEUS (DIS), took the values for all the branching ratios involved
from~\protect\cite{jp:g33:1}, the rest of the quoted measurements took
them from~\protect\cite{pl:b592:1}.
}
\label{tab:gs}
\end{center}
\end{table}
\vspace*{2.cm}
\begin{table}[hbt]
\begin{center}
\begin{tabular}{|c|c|} \hline
& $P^d_{\rm v}$ \\
\hline
\hline
ZEUS (DIS)
& $0.617\pm0.038({\rm stat.})^{+0.017}_{-0.009}({\rm syst.})\pm0.017({\rm br.})$ \\
\hline
ZEUS ($\gamma p$)~\cite{epj:c44:351}
& $0.566\pm0.025({\rm stat.})^{+0.007}_{-0.022}({\rm syst.})^{+0.022}_{-0.023}({\rm br.})$ \\
\hline
combined $e^+e^-$ data
& $0.614\pm0.019({\rm stat.\oplus syst.})^{+0.023}_{-0.025}({\rm br.})$ \\
\hline
H1 (DIS)~\cite{epj:c38:447}
& $0.693\pm0.045({\rm stat.})\pm0.004({\rm syst.})\pm0.009({\rm br.\oplus theory})$ \\
\hline
\end{tabular}
\caption{
The fraction of charged $D$ mesons produced in a vector state, $P^d_{\rm v}$.
The $e^+e^{-}$ values are taken from ~\protect\cite{epj:c44:351}; they are an
update of the compilation in~\protect\cite{hep-ex-9912064p} using the
branching-ratio values of~\protect\cite{pl:b592:1}. The measurements in this
paper, ZEUS (DIS), took the values for all the branching ratios involved
from~\protect\cite{jp:g33:1}, the rest of the quoted measurements took
them from~\protect\cite{pl:b592:1}.
}
\label{tab:pv}
\end{center}
\end{table}
\newpage 
\clearpage
\begin{landscape}
\begin{table}[hbt]
\begin{center}
\begin{tabular}{|c|c|c|c|c|} \hline
& ZEUS (DIS) & ZEUS ($\gamma p$)~\cite{epj:c44:351} & Combined & H1 (DIS) \\
& $p_T(D)>3\,$GeV & $p_T(D)>3.8\,$GeV & $e^+e^-$ data~\cite{hep-ex-9912064p} & \cite{epj:c38:447} \\
& $|\eta(D)|<1.6 $& $|\eta(D)|<1.6$ &  & \\
\hline
\hline
&\phantom{~~~~~~~~~} stat.\phantom{~~} syst.\phantom{~} br.
&\phantom{~~~~~~~~~} stat.\phantom{~~} syst.
&\phantom{~~~~}stat.$\oplus\,$syst.\phantom{~} br.
&\phantom{~~~~~~~~~~}total \\
\hline
$\fcdc$ & 
$0.216 \pm 0.019\phantom{~}^{+0.002\, +0.008}_{-0.020\, -0.010}$  &
$0.217 \pm 0.014\phantom{~}^{+0.013}_{-0.005}$  &
$0.226\phantom{~} \pm 0.010\phantom{~~} ^{+0.016}_{-0.014}$ & $0.203 \pm 0.026$ \\
\hline
$\fcdz$ &
$0.605 \pm 0.020\phantom{~}^{+0.009\, +0.015}_{-0.052\, -0.023}$  &
$0.523 \pm 0.021\phantom{~}^{+0.018}_{-0.017}$  &
$0.557\phantom{~} \pm 0.023\phantom{~~} ^{+0.014}_{-0.013}$ & $0.560 \pm 0.046$ \\
\hline
$\fcdss$ & 
$0.092 \pm 0.011\phantom{~}^{+0.007\, +0.012}_{-0.008\, -0.010}$ &
$0.095 \pm 0.008\phantom{~}^{+0.005}_{-0.005}$ &
$0.101\phantom{~} \pm 0.009\phantom{~~} ^{+0.034}_{-0.020}$ & $0.151 \pm 0.055$ \\
\hline
$\fcds$ &
 $0.229 \pm 0.011\phantom{~}^{+0.006\, +0.007}_{-0.021\, -0.010}$ &
 $0.200 \pm 0.009\phantom{~}^{+0.008}_{-0.006}$ &
$0.238\phantom{~} \pm 0.007\phantom{~~}^{+0.003}_{-0.003}$ & $0.263 \pm 0.032$ \\
\hline
\end{tabular}
\caption{
The fractions of $c$ quarks hadronising as a particular charm hadron,
$f(c \rightarrow D)$.
The fractions are shown for the $D^+$, $D^0$ and $D^+_s$
charm ground states and for the $D^{*+}$ state.
The measurements in this paper, ZEUS (DIS), took the values for all the
branching ratios involved from~\protect\cite{jp:g33:1},
the rest of the quoted measurements took them from~\protect\cite{pl:b592:1}.
}
\label{tab:ff}
\end{center}
\end{table}

\end{landscape}
\newpage

\begin{table}[p]
\centering
\centerline{\Large $\quad\quad\quad\quad\quad\quad\quad\quad\quad$ untagged $D^0$
$\quad\quad\quad\quad\quad\quad\quad\quad$ $D^+$}
\begin{tabular}{|rl||ccc||ccc|}
\hline
$Q^2$ bin&\hspace*{-0.4cm} ($\gev^2$) & $d\sigma / dQ^2$ & $\Delta_{\rm stat}$ & $\Delta_{\rm syst}$ & $d\sigma / dQ^2$ & $\Delta_{\rm stat}$ & $\Delta_{\rm syst}$ \\
& & & (nb/$\gev^2$) & & & (nb/$\gev^2$) & \\
\hline
1.5,&\hspace*{-0.3cm}5    & $0.56$    & $\pm 0.08$    &   $ ^{+0.09} _{-0.03}$     & $0.275$   & $\pm 0.066$    &   $ ^{+0.056} _{-0.044}$\\
5,&\hspace*{-0.3cm}15     & $0.141$   & $\pm 0.008$   &   $ ^{+0.011} _{-0.006}$   & $0.089$   & $\pm 0.014$    &   $ ^{+0.003} _{-0.010}$\\
15,&\hspace*{-0.3cm}40    & $0.044$   & $\pm 0.005$   &   $ ^{+0.003} _{-0.002}$   & $0.016$   & $\pm 0.003$    &   $ ^{+0.002} _{-0.001}$\\
40,&\hspace*{-0.3cm}1000  & $0.0012$  & $\pm 0.0002$  &   $ ^{+0.0002} _{-0.0001}$ & $0.0007$  & $\pm 0.0002$   &   $ ^{+0.0001} _{-0.0001}$\\

\hline
\hline
$x$&\hspace*{-0.4cm} bin & $d\sigma / dx$ & $\Delta_{\rm stat}$ & $\Delta_{\rm syst}$ & $d\sigma / dx$ & $\Delta_{\rm stat}$ & $\Delta_{\rm syst}$ \\
& & & (nb) & & & (nb) & \\
\hline
0.000021,&\hspace*{-0.3cm} 0.0001 & $15197$ & $\pm 2543$ & $ ^{+1137} _{-5789}$& $4470$ & $\pm 1572$  & $ ^{+1078} _{-1112}$\\
0.0001,&\hspace*{-0.3cm} 0.0005   & $4162$  & $\pm 304$  & $ ^{+457} _{-231} $  & $2295$  & $\pm 327$ &  $ ^{+157} _{-199}$\\
0.0005,&\hspace*{-0.3cm} 0.001    & $1476$  & $\pm 195$  & $ ^{+128} _{-72}  $   & $1049$  & $\pm 178$ &  $ ^{+39} _{-125}$\\
0.001,&\hspace*{-0.3cm} 0.1       & $20.7$  & $\pm 2.3$  & $ ^{+2.7} _{-1.0} $  & $10.5$  & $\pm 2.2$ &  $ ^{+1.4} _{-0.5}$\\
\hline
\hline
$p_T(D)$&\hspace*{-0.4cm} bin ($\gev$) & $d\sigma / dp_T(D^0)$ & $\Delta_{\rm stat}$ & $\Delta_{\rm syst}$ & $d\sigma / dp_T(D^+)$ & $\Delta_{\rm stat}$ & $\Delta_{\rm syst}$ \\
& & & (nb/$\gev$) & & & (nb/$\gev$) & \\
\hline
3.0,&\hspace*{-0.3cm} 3.5  & $3.13$  & $ \pm 0.39$  & $ ^{+0.28} _{-0.12}$   & $1.61$  & $ \pm 0.44$  & $ ^{+0.45} _{-0.21}$\\
3.5,&\hspace*{-0.3cm} 4.5  & $1.93$  & $ \pm 0.20$  & $ ^{+0.12} _{-0.11}$   & $0.77$  & $ \pm 0.14$  & $ ^{+0.09} _{-0.06}$\\
4.5,&\hspace*{-0.3cm} 6.0  & $0.78$  & $ \pm 0.11$  & $ ^{+0.05} _{-0.08}$   & $0.49$  & $ \pm 0.08$  & $ ^{+0.05} _{-0.02}$\\
6.0,&\hspace*{-0.3cm} 20.  & $0.051$ & $ \pm 0.009$ & $ ^{+0.004} _{-0.003}$ & $0.028$ & $ \pm 0.007$ & $ ^{+0.002} _{-0.001}$\\
\hline
\hline
$\eta(D)$&\hspace*{-0.4cm} bin & $d\sigma / d\eta(D^0)$ & $\Delta_{\rm stat}$ & $\Delta_{\rm syst}$ & $d\sigma / d\eta(D^+)$ & $\Delta_{\rm stat}$ & $\Delta_{\rm syst}$ \\
& & & (nb) & & & (nb) & \\
\hline
$-1.6,$&\hspace*{-0.3cm} $-0.6$  & $1.18$ & $ \pm 0.19$ & $ ^{+0.13} _{-0.10}$ & $0.65$ & $ \pm 0.11$ &  $ ^{+0.08} _{-0.09}$\\
$-0.6,$&\hspace*{-0.3cm} 0.0     & $1.59$ & $ \pm 0.19$ & $ ^{+0.10} _{-0.11}$ & $1.20$ & $ \pm 0.25$ &  $ ^{+0.15} _{-0.22}$\\
 0.0,&\hspace*{-0.3cm} 0.6       & $2.05$ & $ \pm 0.22$ & $ ^{+0.18} _{-0.14}$ & $1.06$ & $ \pm 0.21$ &  $ ^{+0.08} _{-0.15}$\\
 0.6,&\hspace*{-0.3cm} 1.6       & $2.31$ & $ \pm 0.37$ & $ ^{+0.09} _{-0.20}$ & $0.74$ & $ \pm 0.23$ &  $ ^{+0.22} _{-0.07}$ \\
\hline
\end{tabular}
\caption{\textit{Measured differential cross sections for $D^0$ not coming from a $D^{*+}$ (left), and $D^+$ (right)
as a function of $Q^2$, $x$,
$p_T(D)$ and $\eta(D)$ for $1.5<Q^2<1000$ GeV$^2$, $0.02<y<0.7$,
\mbox{$p_T(D) > 3$ GeV} and $|\eta(D)|<1.6$.
The estimated $b$-quark contribution of  3.1 \%  has been subtracted.
The statistical and systematic
uncertainties are shown separately.
The $D^0$ ($D^+$) cross sections have a further $1.8\%$ ($3.6\%$) uncertainty
from the
$D^0\rightarrow K^-\pi^+$ ($D^+\rightarrow K^-\pi^+\pi^+$) branching ratios.
}}
\label{tab:d0dpm_bins}
\end{table}

\begin{table}[p]
\centering
\centerline{\Large $D_s^+$}
\begin{tabular}{|rl|ccc|}
\hline
$Q^2$ bin&\hspace*{-0.4cm} ($\gev^2$) & $d\sigma / dQ^2$ & $\Delta_{\rm stat}$ & $\Delta_{\rm syst}$ \\
& & & (nb/$\gev^2$) & \\
\hline
1.5,&\hspace*{-0.3cm} 5    & $0.31$    & $\pm 0.07$    &   $ ^{+0.08} _{-0.05}$  \\
5,&\hspace*{-0.3cm} 15     & $0.092$   & $\pm 0.016$   &   $ ^{+0.004} _{-0.017}$  \\
15,&\hspace*{-0.3cm} 40    & $0.016$   & $\pm 0.005$   &   $ ^{+0.004} _{-0.003}$  \\
40,&\hspace*{-0.3cm} 1000  & $0.00025$  & $\pm 0.00010$  &   $ ^{+0.00008} _{-0.00004}$  \\

\hline
\hline
$x$&\hspace*{-0.4cm} bin & $d\sigma / dx$ & $\Delta_{\rm stat}$ & $\Delta_{\rm syst}$ \\ 
& & & (nb) & \\
\hline
0.000021,&\hspace*{-0.3cm} 0.0001 & $4982$ & $\pm 1967$  & $ ^{+1354} _{-1333}$ \\
0.0001,&\hspace*{-0.3cm} 0.0005   & $2765$  & $\pm 443$  & $ ^{+65} _{-644}$ \\
0.0005,&\hspace*{-0.3cm} 0.001    & $934$  & $\pm 250$  & $ ^{+118} _{-155}$ \\
0.001,&\hspace*{-0.3cm} 0.1       & $ 6.1$  & $\pm 1.5$  & $ ^{+0.8} _{-0.6}$ \\
\hline
\hline
$p_T(D_s^+)$&\hspace*{-0.4cm} bin ($\gev$) & $d\sigma / dp_T(D_s^+)$ & $\Delta_{\rm stat}$ & $\Delta_{\rm syst}$\\ 
& & & (nb/$\gev$) & \\
\hline
2.0,&\hspace*{-0.3cm} 2.5  & $1.65$  & $ \pm 0.52$  & $ ^{+0.36} _{-0.50}$  \\  
2.5,&\hspace*{-0.3cm} 3.0  & $0.62$  & $ \pm 0.22$  & $ ^{+0.14} _{-0.11}$  \\
3.0,&\hspace*{-0.3cm} 3.5  & $0.59$  & $ \pm 0.21$  & $ ^{+0.08} _{-0.12}$  \\
3.5,&\hspace*{-0.3cm} 4.5  & $0.55$  & $ \pm 0.11$  & $ ^{+0.05} _{-0.05}$  \\
4.5,&\hspace*{-0.3cm} 6.0  & $0.20$  & $ \pm 0.05$  & $ ^{+0.02} _{-0.01}$  \\
6.0,&\hspace*{-0.3cm} 20.  & $0.011$ & $ \pm 0.004$ & $ ^{+0.002} _{-0.001}$  \\
\hline
\hline
$\eta(D_s^+)$&\hspace*{-0.4cm} bin & $d\sigma / d\eta(D_s^+)$ & $\Delta_{\rm stat}$ & $\Delta_{\rm syst}$ \\ 
& & & (nb) & \\
\hline
$-1.6$,&\hspace*{-0.3cm} $-0.6$  & $0.94$ & $ \pm 0.24$ & $ ^{+0.11} _{-0.26}$ \\
$-0.6$,&\hspace*{-0.3cm} 0.0     & $0.57$ & $ \pm 0.15$ & $ ^{+0.14} _{-0.04}$ \\ 
 0.0,&\hspace*{-0.3cm} 0.6       & $0.76$ & $ \pm 0.18$ & $ ^{+0.06} _{-0.09}$ \\
 0.6,&\hspace*{-0.3cm} 1.6       & $0.85$ & $ \pm 0.22$ & $ ^{+0.17} _{-0.12}$ \\
\hline
\end{tabular}
\caption{\textit{Measured  $D_s^+$  differential cross sections as a function of $Q^2$, $x$,
$p_T(D_s^+)$ and $\eta(D_s^+)$ for $1.5<Q^2<1000$ GeV$^2$, $0.02<y<0.7$,
\mbox{$p_T(D_s^+)> 2$ GeV} and $|\eta(D_s^+)|<1.6$.
The estimated $b$-quark contribution of  4.3 \%  has been subtracted.
The statistical and systematic uncertainties are shown separately.
The cross sections have a further $13\%$ uncertainty from the
$D_s^+\rightarrow  \phi\pi^+ \rightarrow K^+K^-\pi^+$ branching ratio.
}}
\label{tab:ds_bins}
\end{table}

\newpage
\begin{table}[hbt]
\begin{center}
\begin{tabular}{|c||c||c||c|c|c|c|c|c|c|c|} \hline
& $\delta_1-\delta_9$
& $\delta_1-\delta_8$
& $\delta_1$
& $\delta_2$
& $\delta_3$
& $\delta_4$
& $\delta_5$
& $\delta_6$
& $\delta_7$
& $\delta_8$
\\
& $(\%)$
& $(\%)$
& $(\%)$
& $(\%)$
& $(\%)$
& $(\%)$
& $(\%)$
& $(\%)$
& $(\%)$
& $(\%)$
\\
\hline
\hline
$\sigma^{\rm untag}(\dz)$
&
$^{+5.8}_{-4.7}$
&
$^{+5.4}_{-4.1}$
&
$^{+2.5}_{-0.0}$
&
$^{+1.7}_{-3.3}$
&
$^{+1.3}_{-0.9}$
&
$^{+1.5}_{-0.6}$
&
$^{+3.5}_{-0.4}$
&
$^{+2.2}_{-2.1}$
&
&
\\
\hline
$\sigma(D^+)$
&
$^{+6.6}_{-5.0}$
&
$^{+3.2}_{-4.6}$
&
$^{+0.6}_{-0.0}$
&
$^{+1.5}_{-3.1}$
&
$^{+1.0}_{-1.9}$
&
$^{+1.4}_{-0.6}$
&
$^{+0.0}_{-1.5}$
&
$^{+2.3}_{-2.2}$
&
&
\\
\hline
$\sigma_2(D_s^+)$
& 
$^{+9.1}_{-7.4}$
& 
$^{+8.8}_{-7.2}$
& 
$^{+0.0}_{-2.0}$
& 
$^{+0.0}_{-4.0}$
& 
$^{+1.2}_{-0.0}$
& 
$^{+0.5}_{-0.4}$
& 
$^{+8.0}_{-5.1}$
& 
$^{+2.2}_{-2.1}$
&
&
\\
\hline
$\sigma^{\rm tag}(\dz)$
&
&
$^{+5.6}_{-4.6}$
&
$^{+0.0}_{-1.5}$
&
$^{+1.8}_{-3.5}$
&
$^{+2.3}_{-1.1}$
&
$^{+1.5}_{-0.3}$
&
$^{+4.0}_{-0.8}$
&
$^{+2.2}_{-2.1}$
&
&
\\
\hline
$\sigma^{\rm add}(D^{*+})$
&
&
$^{+9.9}_{-8.9}$
&
$^{+0.0}_{-2.9}$
&
$^{+1.9}_{-3.8}$
&
$^{+4.2}_{-0.2}$
&
$^{+2.3}_{-0.5}$
&
$^{+8.2}_{-7.2}$
&
$^{+2.2}_{-2.1}$
&
&
\\
\hline
$\sigma^{\rm kin}(D^{*+})$
&
&
$^{+5.7}_{-4.7}$
&
$^{+0.0}_{-1.8}$
&
$^{+1.8}_{-3.6}$
&
$^{+2.5}_{-0.8}$
&
$^{+1.6}_{-0.3}$
&
$^{+4.0}_{-1.1}$
&
$^{+2.2}_{-2.1}$
&
&
\\
\hline
$\sigma(D_s^+)$
&
&
$^{+8.9}_{-4.9}$
&
$^{+2.8}_{-0.0}$
&
$^{+2.2}_{-4.5}$
&
$^{+4.0}_{-0.0}$
&
$^{+0.4}_{-0.1}$
&
$^{+6.8}_{-0.0}$
&
$^{+2.2}_{-2.1}$
&
&
\\
\hline
$R_{u/d}$
&
&
$^{+4.3}_{-1.4}$
&
$^{+2.7}_{-0.0}$
&
$^{+0.0}_{-0.1}$
&
$^{+1.4}_{-1.4}$
&
$^{+0.1}_{-0.1}$
&
$^{+3.0}_{-0.3}$
&
$^{+0.1}_{-0.1}$
&
&
\\
\hline
$\gamma_s$
&
&
$^{+7.9}_{-3.0}$
&
$^{+1.8}_{-0.0}$
&
$^{+0.5}_{-1.2}$
&
$^{+4.1}_{-0.4}$
&
$^{+0.6}_{-1.3}$
&
$^{+6.4}_{-2.4}$
&
$^{+0.0}_{-0.0}$
&
&
\\
\hline
$P^d_{\rm v}$
&
&
$^{+2.8}_{-1.4}$
&
$^{+0.0}_{-1.3}$
&
$^{+0.1}_{-0.3}$
&
$^{+1.4}_{-0.0}$
&
$^{+0.2}_{-0.0}$
&
$^{+2.4}_{-0.0}$
&
$^{+0.0}_{-0.1}$
&
&
\\
\hline
$\sigma_{\rm gs}$
&
&
$^{+10.3}_{-4.3}$
&
$^{+1.2}_{-0.0}$
&
$^{+1.7}_{-3.4}$
&
$^{+1.2}_{-0.5}$
&
$^{+1.4}_{-0.5}$
&
$^{+2.4}_{-0.5}$
&
$^{+2.2}_{-2.1}$
&
$^{+9.3}_{-0.9}$
&
$^{+0.4}_{-0.4}$
\\
\hline
$\fcdc$
&
&
$^{+1.1}_{-9.2}$
&
$^{+0.0}_{-0.7}$
&
$^{+0.3}_{-0.2}$
&
$^{+0.1}_{-1.7}$
&
$^{+0.1}_{-0.2}$
&
$^{+0.2}_{-2.9}$
&
$^{+0.1}_{-0.1}$
&
$^{+0.9}_{-8.5}$
&
$^{+0.4}_{-0.4}$
\\
\hline
$\fcdz$
&
&
$^{+1.1}_{-8.6}$
&
$^{+0.0}_{-0.0}$
&
$^{+0.0}_{-0.0}$
&
$^{+0.5}_{-0.4}$
&
$^{+0.2}_{-0.1}$
&
$^{+1.8}_{-0.0}$
&
$^{+0.0}_{-0.0}$
&
$^{+0.9}_{-8.5}$
&
$^{+0.4}_{-0.4}$
\\
\hline
$f^{\rm untag}(c\rightarrow D^0)$
&
&
$^{+2.0}_{-8.6}$
&
$^{+1.0}_{-0.0}$
&
$^{+0.1}_{-0.0}$
&
$^{+0.5}_{-0.9}$
&
$^{+0.2}_{-0.1}$
&
$^{+1.4}_{-0.2}$
&
$^{+0.0}_{-0.0}$
&
$^{+0.9}_{-8.5}$
&
$^{+0.4}_{-0.4}$
\\
\hline
$\fcdss$
&
&
$^{+7.1}_{-8.9}$
&
$^{+1.6}_{-0.0}$
&
$^{+0.5}_{-1.0}$
&
$^{+3.7}_{-0.3}$
&
$^{+0.5}_{-1.2}$
&
$^{+5.7}_{-2.1}$
&
$^{+0.0}_{-0.0}$
&
$^{+0.9}_{-8.5}$
&
$^{+0.4}_{-0.4}$
\\
\hline
$\fcds$
&
&
$^{+2.8}_{-9.1}$
&
$^{+0.0}_{-2.9}$
&
$^{+0.1}_{-0.1}$
&
$^{+1.5}_{-0.3}$
&
$^{+0.2}_{-0.0}$
&
$^{+2.2}_{-1.5}$
&
$^{+0.0}_{-0.0}$
&
$^{+0.9}_{-8.5}$
&
$^{+0.4}_{-0.4}$
\\
\hline
\end{tabular}
\caption{
The systematic uncertainties resulting from $\delta_1$-$\delta_9$,
from $\delta_1$-$\delta_8$, and from $\delta_1$ to $\delta_8$
independently (see text) for the charm hadron
cross sections and charm fragmentation ratios and fractions.
}
\label{tab:syst}
\end{center}
\end{table}

\newpage

\begin{table}[htb]
\begin{center}
\centerline{\Large untagged $D^0$}
\vspace*{0.1cm}
\begin{tabular}{|rl|c|cccc|}
\hline
$Q^2$ bin&\hspace*{-0.4cm} $(\gev^2)$ & $y$ bin  & $\sigma$ & $\Delta_{\rm stat}$ & $\Delta_{\rm syst}$ & (nb) \\
\hline \hline
1.5,&\hspace*{-0.3cm} 9    & 0.18, 0.70 & $1.50$ & $ \pm 0.19$ & $ ^{+0.14} _{-0.16} $ &  \\
 &          & 0.02, 0.18 & $1.03$ & $ \pm 0.19$ & $ ^{+0.16} _{-0.12} $ &  \\
\hline
9,&\hspace*{-0.3cm} 44   & 0.20, 0.70 & $1.02$ & $ \pm 0.14$ & $ ^{+0.12} _{-0.09} $ &  \\
  &       & 0.02, 0.20 & $1.02$ & $ \pm 0.14$ & $ ^{+0.12} _{-0.05} $ &  \\
\hline
44,&\hspace*{-0.3cm} 1000 & 0.02, 0.70 & $1.03$ & $ \pm 0.19$ & $ ^{+0.16} _{-0.05} $ &  \\
\hline
\end{tabular}
\newline
\vspace*{0.5cm}
\centerline{\Large $D_{\quad}^+$}
\vspace*{0.1cm}
\begin{tabular}{|rl|c|cccc|}
\hline
$Q^2$ bin&\hspace*{-0.4cm} $(\gev^2)$ & $y$ bin  & $\sigma$ & $\Delta_{\rm stat}$ & $\Delta_{\rm syst}$ & (nb) \\
\hline \hline
1.5,&\hspace*{-0.3cm} 9    & 0.18, 0.70 & $0.63$ & $ \pm 0.14$ & $ ^{+0.05} _{-0.10} $ &  \\
   &        & 0.02, 0.18 & $0.65$ & $ \pm 0.13$ & $ ^{+0.09} _{-0.08} $ &  \\
\hline
9,&\hspace*{-0.3cm} 44   & 0.20, 0.70 & $0.52$ & $ \pm 0.11$ & $ ^{+0.03} _{-0.10} $ &  \\
    &        & 0.02, 0.20 & $0.44$ & $ \pm 0.11$ & $ ^{+0.06} _{-0.03} $ &  \\
\hline
44,&\hspace*{-0.3cm} 1000   & 0.02, 0.70 & $0.61$ & $ \pm 0.25$ & $ ^{+0.08} _{-0.09} $ &  \\
\hline
\end{tabular}
\newline
\vspace*{0.5cm}
\centerline{\Large $D_{s}^+$}
\vspace*{0.1cm}
\begin{tabular}{|rl|c|cccc|}
\hline
$Q^2$ bin&\hspace*{-0.4cm} $(\gev^2)$ & $y$ bin  & $\sigma$ & $\Delta_{\rm stat}$ & $\Delta_{\rm syst}$ & (nb) \\
\hline \hline
1.5,&\hspace*{-0.3cm} 9    & 0.18, 0.70 & $0.97$ & $ \pm 0.23$ & $ ^{+0.16} _{-0.21} $ &  \\
    &     & 0.02, 0.18 & $0.56$ & $ \pm 0.15$ & $ ^{+0.11} _{-0.08} $ &  \\
\hline
9,&\hspace*{-0.3cm} 44   & 0.20, 0.70 & $0.61$ & $ \pm 0.15$ & $ ^{+0.14} _{-0.15} $ &  \\
    &   & 0.02, 0.20 & $0.29$ & $ \pm 0.07$ & $ ^{+0.12} _{-0.03} $ &  \\
\hline
44,&\hspace*{-0.3cm} 1000   & 0.02, 0.70 & $0.20$ & $ \pm 0.11$ & $ ^{+0.08} _{-0.03} $ &  \\
\hline
\end{tabular}
\newline
\end{center}
\caption{\textit{Measured cross sections for $D^0$ not coming from a $D^{*+}$,
$D^+$ and $D_s^+$ in each of the $Q^2$ and $y$ bins for
$p_T(D^0,D^+) > 3$ GeV,  $p_T(D_s^+) > 2$ GeV and $|\eta(D)|<1.6$.
The estimated $b$-quark contribution of 3.1 \%  for $D^0$
and $D^+$ and  4.3 \%  for $D_s^+$ has been subtracted.
The statistical and systematic uncertainties are shown separately.
The $D^0$, $D^+$ and $D_s^+$ cross sections have further $1.8\%$, $3.6\%$ and
$13\%$ uncertainties from the
$D^0\rightarrow K^-\pi^+$, $D^+\rightarrow K^-\pi^+\pi^+$ and
$D_s^+\rightarrow  \phi\pi^+ \rightarrow K^+K^-\pi^+$ branching ratios,
respectively.}}
\label{tab:dd_bins}
\end{table}

\newpage
\clearpage

\begin{table}[htb]
\begin{center}
\vspace*{-1.0cm}
\centerline{\Large untagged $D^0$}
\vspace*{0.1cm}
\begin{tabular}{|c|c|ccc|cc|}
\hline
$Q^2$ ($\gev^2$) & $x$ & $F_2^{c\bar{c}}$ & $\Delta_{\rm stat}$ & $\Delta_{\rm syst}$
& $\Delta_{\rm extrap}$ & factor \\
\hline \hline
4.2 & 0.00013 & $0.141$ & $ \pm 0.017$ & $ ^{+0.013} _{-0.015}$ & $ ^{+0.048} _{-0.013}$ &  8.9 \\
    & 0.00061 & $0.090$ & $ \pm 0.017$ & $ ^{+0.014} _{-0.011}$ & $ ^{+0.036} _{-0.006}$ &  17 \\
\hline
20.4& 0.00062 & $0.320$ & $ \pm 0.044$ & $ ^{+0.037} _{-0.029}$ & $ ^{+0.061} _{-0.020}$ &  4.9 \\
    & 0.00281 & $0.156$ & $ \pm 0.021$ & $ ^{+0.019} _{-0.008}$ & $ ^{+0.041} _{-0.004}$ &  5.8 \\
\hline 
111.8 & 0.00676 & $0.217$ & $ \pm 0.039$ & $ ^{+0.033} _{-0.011}$ & $ ^{+0.014} _{-0.002}$ &  2.5 \\
\hline
\end{tabular}
\newline
\vspace*{0.2cm}
\centerline{\Large $D_{\quad}^+$}
\vspace*{0.1cm}
\begin{tabular}{|c|c|ccc|cc|}
\hline
$Q^2$ ($\gev^2$) & $x$ & $F_2^{c\bar{c}}$ & $\Delta_{\rm stat}$ & $\Delta_{\rm syst}$
& $\Delta_{\rm extrap}$ & factor \\
\hline \hline
4.2 & 0.00013 & $0.123$ & $ \pm 0.025$ & $ ^{+0.010} _{-0.020}$ & $ ^{+0.037} _{-0.011}$ &  8.9 \\
    & 0.00061 & $0.109$ & $ \pm 0.020$ & $ ^{+0.015} _{-0.014}$ & $ ^{+0.039} _{-0.007}$ &  17 \\
\hline
20.4& 0.00062 & $0.331$ & $ \pm 0.067$ & $ ^{+0.016} _{-0.067}$ & $ ^{+0.066} _{-0.021}$ &  4.9 \\
    & 0.00281 & $0.130$ & $ \pm 0.039$ & $ ^{+0.017} _{-0.009}$ & $ ^{+0.030} _{-0.003}$ &  5.8 \\
\hline
111.8 & 0.00676 & $0.293$ & $ \pm 0.124$ & $ ^{+0.037} _{-0.041}$ & $ ^{+0.021} _{-0.003}$ &  2.5 \\
\hline
\end{tabular}
\newline
\vspace*{0.2cm}
\centerline{\Large $D_s^+$} 
\vspace*{0.1cm}
\begin{tabular}{|c|c|ccc|cc|}
\hline
$Q^2$ ($\gev^2$) & $x$ & $F_2^{c\bar{c}}$ & $\Delta_{\rm stat}$ & $\Delta_{\rm syst}$
& $\Delta_{\rm extrap}$ & factor \\
\hline \hline
4.2 & 0.00013 & $0.221$ & $ \pm 0.044$ & $ ^{+0.037} _{-0.048}$ & $ ^{+0.036} _{-0.016}$ &  4.3 \\
    & 0.00061 & $0.075$ & $ \pm 0.017$ & $ ^{+0.016} _{-0.011}$ & $ ^{+0.019} _{-0.004}$ &  5.6 \\
\hline
20.4& 0.00062 & $0.470$ & $ \pm 0.100$ & $ ^{+0.109} _{-0.112}$ & $ ^{+0.037} _{-0.017}$ &  2.8 \\
    & 0.00281 & $0.100$ & $ \pm 0.022$ & $ ^{+0.043} _{-0.009}$ & $ ^{+0.013} _{-0.001}$ &  2.9 \\
\hline
111.8& 0.00676 & $0.179$ & $ \pm 0.058$ & $ ^{+0.075} _{-0.025}$ & $ ^{+0.013} _{-0.001}$ &  1.9 \\
\hline
\end{tabular}
\newline
\vspace*{0.2cm}
\centerline{\Large Combined}
\vspace*{0.1cm}
\begin{tabular}{|c|c|ccc|c|}
\hline
$Q^2$ ($\gev^2$) & $x$ & $F_2^{c\bar{c}}$ & $\Delta_{\rm stat}$ & $\Delta_{\rm syst}$
& $\Delta_{\rm extrap}$ \\
\hline \hline
4.2 & 0.00013 & $0.144$ & $ \pm 0.014$ & $ ^{+0.022} _{-0.015}$ & $ ^{+0.045} _{-0.013}$  \\
    & 0.00061 & $0.090$ & $ \pm 0.010$ & $ ^{+0.010} _{-0.004}$ & $ ^{+0.029} _{-0.005}$  \\
\hline
20.4& 0.00062 & $0.341$ & $ \pm 0.035$ & $ ^{+0.046} _{-0.042}$ & $ ^{+0.063} _{-0.021}$  \\
    & 0.00281 & $0.132$ & $ \pm 0.014$ & $ ^{+0.024} _{-0.005}$ & $ ^{+0.024} _{-0.001}$  \\
\hline
111.8 & 0.00676 & $0.211$ & $ \pm 0.032$ & $ ^{+0.044} _{-0.013}$ & $ ^{+0.013} _{-0.002}$  \\
\hline
\end{tabular}
\caption{\textit{
The extracted values of $F_2^{c\bar{c}}$ from the production cross sections of
$D^0$ not coming from $D^{*+}$, $D^+$ and $D_s^+$ and the combination of them
at each $Q^2$ and $x$ value.
The statistical, systematic and extrapolation uncertainties are shown separately.
The values of the extrapolation factor used to correct the full
$p_T(D)$  and $\eta(D)$ phase space are also shown.
All the extracted $F_2^{c\bar{c}}$ values  have a further $+2.7\%$ $-4.1\%$
uncertainty from the
$D^0\rightarrow K^-\pi^+$, $D^+\rightarrow K^-\pi^+\pi^+$ and
$D_s^+\rightarrow  \phi\pi^+ \rightarrow K^+K^-\pi^+$
branching ratios and the $f(c\rightarrow \Lambda_c^+)$ value.
}}
\label{tab:f2c_bins}
\end{center}
\end{table}

%
\newpage
\newpage
\newpage
\newpage
\newpage

\newpage

%
%
\end{document}

%% file: zeus_def.tex
\newcommand{\ZcoosysB}{%
The ZEUS coordinate system is a right-handed Cartesian system, with the $Z$
axis pointing in the proton beam direction, referred to as the ``forward
direction'', and the $X$ axis pointing left towards the centre of HERA.
The coordinate origin is at the nominal interaction point.\xspace}

\newcommand{\ZcoosysfnB}{\footnote{\ZcoosysB}}

\newcommand{\Zdetdesc}{%
A detailed description of the ZEUS detector can be found 
elsewhere~\cite{zeus:1993:bluebook}. A brief outline of the 
components that are most relevant for this analysis is given
below.\xspace}
\newcommand{\Zctddesc}[1]{%
Charged particles are tracked in the central tracking detector (CTD)~\citeCTD,
which operates in a magnetic field of $1.43\Tesla$ provided by a thin 
superconducting coil. The CTD consists of 72~cylindrical drift chamber 
layers, organised in 9~superlayers covering the polar-angle#1 region 
\mbox{$15^\circ<\theta<164^\circ$}. The transverse-momentum resolution for
full-length tracks is $\sigma(p_T)/p_T=0.0058p_T\oplus0.0065\oplus0.0014/p_T$,
with $p_T$ in $\Gev$.}
\newcommand{\Zcaldesc}{%
The high-resolution uranium--scintillator calorimeter (CAL)~\citeCAL consists 
of three parts: the forward (FCAL), the barrel (BCAL) and the rear (RCAL)
calorimeters. Each part is subdivided transversely into towers and
longitudinally into one electromagnetic section and either one (in RCAL)
or two (in BCAL and FCAL) hadronic sections. The smallest subdivision of
the calorimeter is called a cell.  The CAL energy resolutions, as measured under
test-beam conditions, are $\sigma(E)/E=0.18/\sqrt{E}$ for electrons and
$\sigma(E)/E=0.35/\sqrt{E}$ for hadrons, with $E$ in $\Gev$.}

\chardef\usc=95
\chardef\til=126
\catcode`\@=11 
\DeclareRobustCommand\xdotspace{\futurelet\@let@token\@xdotspace}
\def\@xdotspace{%
  \ifx\@let@token.\else
  \ifx\@let@token\bgroup.\else
  \ifx\@let@token\egroup.\else
  \ifx\@let@token\/.\else
  \ifx\@let@token\ .\else
  \ifx\@let@token~.\else
  \ifx\@let@token!.\else
  \ifx\@let@token,.\else
  \ifx\@let@token:.\else
  \ifx\@let@token;.\else
  \ifx\@let@token?.\else
  \ifx\@let@token/.\else
  \ifx\@let@token'.\else
  \ifx\@let@token).\else
  \ifx\@let@token-.\else
  \ifx\@let@token\@xobeysp.\else
  \ifx\@let@token\space.\else
  \ifx\@let@token\@sptoken.\else
   .\space
   \fi\fi\fi\fi\fi\fi\fi\fi\fi\fi\fi\fi\fi\fi\fi\fi\fi\fi}
\catcode`\@=12 

\newcommand{\stru}[2]{%
   \relax\ifmmode\hbox{\vrule height#1 depth#2 width0pt}%
   \else\vrule height#1 depth#2 width0pt\fi}

\newcommand{\Ronum}[1]{\uppercase\expandafter{\romannumeral#1}}
\newcommand{\ronum}[1]{\expandafter{\romannumeral#1}}
\DeclareRobustCommand{\LaTeXZ}{%
  \LaTeX\kern-.05em4\kern-.1em
  {\raisebox{-0.2ex}{$\scriptstyle\text{ZEUS}$}}\xspace}

\DeclareMathAlphabet{\mathbf}{OT1}{cmr}{bx}{sl}
\newcommand{\eVdist}{\kern-0.06667em}

\newcommand{\Gev}{{\text{Ge}\eVdist\text{V\/}}}

\newcommand{\gev}{{\,\text{Ge}\eVdist\text{V\/}}}

\newcommand{\Tesla}{\,\text{T}}

\newcommand{\slashfrac}[2]{%
  \raisebox{0.5ex}{\ensuremath #1}\kern-0.12em/\kern-0.08em
  \raisebox{-.8ex}{\ensuremath #2}}

\newcommand{\sqr}[3]{%
    {\vcenter{\hrule height.#3ex\hbox{\vrule width.#2ex height#1ex
     \kern#1ex\vrule width.#3ex}\hrule height.#2ex}}}

\catcode`\@=11 
\newcommand{\parenbar}{\mathpalette\p@renb@r}
\def\p@renb@r#1#2{\vbox{%
  \ifx#1\scriptscriptstyle \dimen@.7em\dimen@ii.2em\else
  \ifx#1\scriptstyle \dimen@.8em\dimen@ii.25em\else
  \dimen@1em\dimen@ii.4em\fi\fi \offinterlineskip
  \ialign{\hfill##\hfill\cr
    \vbox{\hrule width\dimen@ii}\cr
    \noalign{\vskip-.3ex}%
    \hbox to\dimen@{$\mathchar300\hfil\mathchar301$}\cr
    \noalign{\vskip-.3ex}%
    $#1#2$\cr}}}
\catcode`\@=12 

\newcommand{\IP}{{\rm I$\kern-0.01667em$P}\xspace}

\mathchardef\qsm=63
\mathchardef\pls=43
\mathchardef\mns=512
\mathchardef\plm=518
\mathchardef\eql=61
\mathchardef\smallleft=300
\mathchardef\smallright=301
\mathchardef\les=316
\mathchardef\gre=318
\mathchardef\leq=532
\mathchardef\grq=533

\catcode`\@=11 
\newcounter{pict@width}
\newcounter{pict@height}
\newlength{\pict@scale}
\newcommand{\psfigadd}[4]{%
\setcounter{pict@width}{1*\ratio{#2+\pict@scale/2}{\pict@scale}}
\setcounter{pict@height}{1*\ratio{#3+\pict@scale/2}{\pict@scale}}
\setlength{\unitlength}{\pict@scale}
\hbox to #2{\hspace{-\fill}\begin{picture}(\thepict@width,\thepict@height)
\put(0,0){\psfig{figure=#1,width=#2,height=#3,clip=}}
\SetScale{0.283466457}
\SetWidth{1.763889}
{#4}
\end{picture}}
}
\newcounter{pict@widthfst}
\newcounter{pict@widthscd}
\newcounter{pict@widthtot}
\newcommand{\psfigaddtwo}[7]{%
\setcounter{pict@widthfst}{1*\ratio{#2+\pict@scale/2}{\pict@scale}}
\setcounter{pict@widthscd}{1*\ratio{#2+#4+\pict@scale/2}{\pict@scale}}
\setcounter{pict@widthtot}{1*\ratio{#2+#4+#6+\pict@scale/2}{\pict@scale}}
\setcounter{pict@height}{1*\ratio{#3+\pict@scale/2}{\pict@scale}}
\setlength{\unitlength}{\pict@scale}
\hbox{\hspace{-\fill}\begin{picture}(\thepict@widthtot,\thepict@height)
\put(0,0){\psfig{figure=#1,width=#2,height=#3,clip=}}
\put(\thepict@widthscd,0){\psfig{figure=#5,width=#6,height=#3,clip=}}
\SetScale{0.283466457}
\SetWidth{1.763889}
{#7}
\end{picture}}
}
\newcommand{\psfigror}[4]{%
\setcounter{pict@width}{1*\ratio{#2+\pict@scale/2}{\pict@scale}}
\setcounter{pict@height}{1*\ratio{#3+\pict@scale/2}{\pict@scale}}
\setlength{\unitlength}{\pict@scale}
\hbox{\begin{picture}(\thepict@width,\thepict@height)
\put(0,\thepict@height){\psfig{figure=#1,width=#3,height=#2,clip=,angle=270}}
\SetScale{0.283466457}
\SetWidth{1.763889}
{#4}
\end{picture}}
}
\newcommand{\psfigrol}[4]{%
\setcounter{pict@width}{1*\ratio{#2+\pict@scale/2}{\pict@scale}}
\setcounter{pict@height}{1*\ratio{#3+\pict@scale/2}{\pict@scale}}
\setlength{\unitlength}{\pict@scale}
\hbox{\begin{picture}(\thepict@width,\thepict@height)
\put(0,0){\psfig{figure=#1,width=#3,height=#2,clip=,angle=90}}
\SetScale{0.283466457}
\SetWidth{1.763889}
{#4}
\end{picture}}
}
\catcode`\@=12 
\newlength\listtextwidth

\catcode`\@=11 
\newlength{\@tabfninsert}
\newlength{\@tabfnwidth}
\newcommand{\tabfootnote}[2]{%
  \setlength{\@tabfninsert}{0.8em}
  \setlength{\@tabfnwidth}{\textwidth}
  \addtolength{\@tabfnwidth}{-\@tabfninsert}
  \addtolength{\@tabfnwidth}{-0.4em}
  \noindent\makebox[\@tabfninsert][r]{\footnotesize$^{#1}$\hfil}\hfill%
  \parbox[t]{\@tabfnwidth}{\footnotesize #2\hfill}}
\catcode`\@=12 

%% file: charm_com.tex
\newcommand{\dsp}        {\mbox{$D^{\ast +}$}}

\newcommand{\dssp}       {\mbox{$D_s^+$}}

\newcommand{\dz}         {\mbox{$D^{0}$}}
\newcommand{\dc}         {\mbox{$D^+$}}

\newcommand{\lcp}        {\mbox{$\Lambda_c^+$}}

\newcommand{\fcdz}       {\mbox{$f(c \rightarrow D^0)$}}
\newcommand{\fcdc}       {\mbox{$f(c \rightarrow D^+)$}}
\newcommand{\fcdss}      {\mbox{$f(c \rightarrow D_s^+)$}}

\newcommand{\fcds}       {\mbox{$f(c \rightarrow D^{\ast +})$}}
\newcommand{\br}         {\mbox{${\cal B}_{D^{\ast +}\rightarrow D^0 \pi^+}$}}
\def\dsk3pi{ {\dsp}~\rightarrow~\dz~\pi^{+}_{s}%
        \rightarrow~(K^{-}~\pi^{+}~\pi^{+}~\pi^{-})~\pi^{+}_{s} }
\def\et10t{ E_T^{\theta > 10^\circ}}
\def\etw10{ E_T^{\theta > 10}}

%% file: physics_com.tex
\newcommand{\qsq}       {\mbox{${Q^2}$}}